\documentclass[a4paper,12pt]{article}
\pdfoutput=1
\usepackage{jheppub_X} 
\usepackage[dvipsnames]{xcolor}
\usepackage{tikz}
\usetikzlibrary{patterns,arrows,decorations.pathmorphing,decorations.markings}

\usepackage{slashed,adjustbox,bm,mathrsfs,bbm}
\usepackage{subcaption}
\usepackage{cleveref}
\usepackage[T1]{fontenc} 
\usepackage[utf8]{inputenc}
\usepackage{mathtools}
\usepackage{float}
\usepackage{amsmath}
\allowdisplaybreaks
\usepackage{setspace}

\usepackage{tikz,environ}
\usepackage{tkz-euclide}
\usetikzlibrary{decorations}
\usetikzlibrary{arrows}
\usetikzlibrary{decorations.markings}
\usetikzlibrary{shapes.geometric}
\tikzset{
	hard/.style={postaction={decorate},
		line width=0.5mm
	},
		soft/.style={postaction={decorate},
		line width=0.5mm, dashed
	},
	momentum/.style={postaction={decorate},
	line width=0.5mm,
	color=gray,
	decoration={
    markings,
    mark=at position 0.8 with {\arrow{stealth}}}
    },
	hardarrow/.style={postaction={decorate},
	line width=0.5mm,
	decoration={
    markings,
    mark=at position 0.6 with {\arrow{stealth}}}},
    triangle/.style={
        draw,
        shape border rotate=90,
        isosceles triangle,
        isosceles triangle apex angle=60,
        node distance=1cm,
        minimum height=2em
    }
}

\crefname{table}{Table}{Tables}
\crefname{equation}{eq.}{eqs.}
\crefname{appendix}{app.}{apps.}
\crefname{section}{sec.}{secs.}
\crefname{figure}{fig.}{figs.}

\newcommand{\nn}{\nonumber \\}
\newcommand{\Lagr}{\mathcal{L}}
\newcommand{\cA}{\mathcal{A}}
\newcommand{\intA}{\mathrm{A}}

\newcommand{\s}{\hspace{0.8pt}}

\newcommand{\AH}[1]{{\color{Mulberry} [AH: #1]}}
\definecolor{colorTC}{rgb}{.2,.7,.2}
\newcommand{\TC}[1]{{\bf \color{colorTC}{[TC: #1]}}}

\newcommand{\FN}[1]{{ \color{red}{[FN: #1]}}}

\title{Scalar soft theorems at one loop}

\preprint{CERN-TH-2025-081}

\title{Geometry of soft scalars at one loop}
\author[a,b,c]{Timothy Cohen,}
\author[d,b]{Ipak Fadakar,}
\author[a]{Andreas Helset,}
\author[b]{and Filippo Nardi\,}

\affiliation[a]{Theoretical Physics Department, CERN, 1211 Geneva, Switzerland}
\affiliation[b]{Theoretical Particle Physics Laboratory, EPFL, 1015 Lausanne, Switzerland}
\affiliation[c]{Institute for Fundamental Science, University of Oregon, Eugene, OR 97403, USA}
\affiliation[d]{Scuola Normale Superiore, Piazza dei Cavalieri 7, Pisa, 56126, Italy}

\emailAdd{tim.cohen@cern.ch}
\emailAdd{ipak.fadakar@sns.it}
\emailAdd{andreas.helset@cern.ch}
\emailAdd{filippo.nardi@epfl.ch}

\abstract{We extend the soft theorems for scattering amplitudes of scalar effective field theories to one-loop order. Our analysis requires carefully accounting for the fact that the soft limit is not guaranteed to commute with evaluating IR-divergent loop integrals; new results for the soft limit of general scalar one-loop integrals are presented. The geometric soft theorem remains unmodified for any derivatively-coupled scalar effective field theory, and we conjecture that this statement holds to all orders. In contrast, the soft theorem receives nontrivial corrections in the presence of potential interactions, analogous to the case of non-Abelian gauge theories. We derive the universal leading-order correction to the scalar soft theorem arising from potential interactions at one loop.  Explicit examples are provided that illustrate the general results.
}

\begin{document}
\maketitle
\flushbottom
\setcounter{page}{2}

\begin{spacing}{1.1}
\parskip=0ex

\section{Introduction}\label{sec:intro}

Universal properties often emerge when studying soft modes in quantum field theory.
The classic examples are soft theorems, which state that there are simple relations among scattering amplitudes for gauge theory \cite{Weinberg:1965nx}, gravity \cite{Weinberg:1965nx}, and pions \cite{Adler:1964um}.  These relations emerge when the energy and momentum of one or more massless modes are sent to zero. Similarly, scattering amplitudes for nonrelativistic Goldstone bosons, such as phonons in superfluids, solids, and fluids, obey universal soft theorems \cite{Green:2022slj,Cheung:2023qwn}.

Soft theorems provide a window into the fundamental properties of the underlying theory. For example, the photon and graviton soft theorems are linked to charge conservation and energy-momentum conservation, respectively \cite{Weinberg:1965nx}. Similarly, the nontrivial double soft theorem for pions exposes their non-Abelian nature \cite{Weinberg:1966kf}. By studying the soft limit of scattering amplitudes, we can uncover universal properties of the theory. This subject has additionally experienced a renaissance as new connections have been made with physics on the celestial sphere; see ref.~\cite{Pasterski:2021raf} for a recent summary.

The leading-order soft theorems at tree level can often be understood to arise from enhancements due to single-particle propagator poles; the soft mode couples to an external hard particle making an internal line go nearly on shell. This is the case for gauge theory and gravity. Whether these simple soft theorems survive at loop level requires detailed study. In fact, for gauge theory we find a variety of answers: the photon soft theorem remains unchanged at loop level \cite{Yennie:1961ad,Grammer:1973db}, while its non-Abelian version picks up nontrivial corrections \cite{Catani:2000pi,Larkoski:2014bxa}. The failure of the tree-level soft theorem for non-Abelian gauge theory is linked to infrared (IR) divergences in the loop integrals \cite{Bern:2014oka}. For gravity, the leading soft theorem is also unchanged at loop level, while subleading soft theorems do receive quantum corrections, see, e.g., refs.~\cite{Bern:2014oka,Beneke:2022pue}.

Recently, a new soft theorem was derived for general scalar effective field theories (EFTs), named the \textit{geometric soft theorem}~\cite{Cheung:2021yog} since it is expressed using field-space geometry. In contrast to traditional soft theorems which are linked to symmetries, the geometric soft theorem is general and applies to generic EFTs.
For scalar theories without potential interactions, it takes the simple form
\begin{align}
    \lim_{q\rightarrow 0}\cA_{n+1} = \nabla_{i} \cA_{n} \,,
    \label{eq:geoSoftThmIntro}
\end{align}
where $\cA_{n}$ is an $n$-particle scattering amplitude and $\nabla_{i}$ is the covariant derivative on field space. In this equation, we are taking the limit $q\rightarrow 0$ for the massless scalar with flavor label $i$. The derivation of the soft theorem  
made no reference to perturbation theory
\cite{Cheung:2021yog}.  However, there is an implicit assumption regarding the order of limits: as long as the soft limit commutes with the evaluation of the loop integral,
the geometric soft theorem should hold without modifications.  This concern was so far not addressed in the literature, and is central to the results presented here.
It is also known that for theories with a potential, the soft theorem is modified. The original derivation of the result with a potential was explicitly restricted to tree level. 

In this paper, we investigate the fate of the geometric soft theorem at one-loop order. First, we consider scalar effective field theories with only derivative couplings.\footnote{We denote a theory to be derivatively-coupled when all interactions involve \textit{some} derivatives, but not necessarily one or more derivative per field. For example, the leading-order EFT for Goldstone bosons---which is derivatively-coupled---has two derivatives for each $n$-point interaction.}  In this case, we will demonstrate that the geometric soft theorem in \cref{eq:geoSoftThmIntro} is unchanged at one loop.  We emphasize that this result makes no reference to symmetry arguments, in contrast with the derivations of the classic soft theorems.
We then turn to theories with potential interactions. 
We will show that when we turn on potential interactions such as $\phi^{3}$, the soft theorem receives one-loop corrections due to the presence of new IR divergences. We will derive the leading quantum corrections to the soft theorem for these theories, and provide a road map for investigating subleading corrections.

The remainder of this paper is organized as follows. In \cref{sec:Background}, we review the geometry of field space and present the tree-level soft theorem for scalar EFTs. Technical details about the soft limit at loop level are summarized in \cref{sec:SoftLimit}. The one-loop extension of the soft theorem is discussed for derivatively-coupled theories in \cref{sec:OneLoopSoftTheorem} and for theories with potential interactions in \cref{sec:1loopPotential}.
In \cref{app:Examples}, we present numerous examples of the soft theorem in action. We conclude in \cref{sec:Conclusion}. In \cref{app:Integrals}, we give details for the one-loop integrals that appear in the examples.

\section{Tree-level soft theorem}\label{sec:Background}
To set the stage, we first review the framework of field-space geometry for EFTs.  The aspect of field-space geometry that we will rely on here is the fact that physical scattering amplitudes for general EFTs can be expressed in terms of geometric quantities such as the Riemann curvature tensor and covariant derivatives thereof.\footnote{Recently, field-space geometry has been successfully applied to Higgs physics \cite{Alonso:2015fsp,Alonso:2016oah,Helset:2018fgq,Corbett:2019cwl,Helset:2020yio,Hays:2020scx,Cohen:2020xca,Corbett:2021eux,Cohen:2021ucp,Alonso:2021rac,Helset:2022pde,Alonso:2022ffe,Assi:2023zid,Jenkins:2023bls,Alonso:2023upf,Li:2024ciy}. There is also a theoretical effort to study the geometric structure of scattering amplitudes \cite{Helset:2022tlf,Helset:2024vle,Lee:2024xqa} and go beyond Riemannian geometry to account for derivative field redefinitions \cite{Finn:2019aip,Cheung:2022vnd,Cohen:2022uuw,Cohen:2023ekv,Craig:2023wni,Craig:2023hhp,Alminawi:2023qtf,Cohen:2024bml,Lee:2024xqa,Aigner:2025xyt}.} Writing the amplitudes geometrically provides a natural way to express the geometric soft theorem for scalar EFTs at tree level \cite{Cheung:2021yog}. We will explore how to extend this soft theorem to one-loop order in the following sections.

\subsection{Geometry of field space}

The scalar soft theorem can be compactly expressed using the language of field-space geometry.  We will restrict ourselves here to EFTs for a set of scalars $\phi^I$, where $I$ is a flavor index.  We write the general EFT Lagrangian as
\begin{align}\label{eq:geoNLSM}
    \Lagr &= \frac{1}{2}\s g_{IJ}(\phi) (\partial_{\mu} \phi^{I}) (\partial^{\mu} \phi^{J}) - V(\phi) \nn[3pt]
    &\hspace{40pt}+ \lambda_{IJKL}(\phi) (\partial_{\mu} \phi^{I})(\partial^{\mu} \phi^{J}) (\partial_{\nu} \phi^{K}) (\partial^{\nu} \phi^{L}) + \dots\,,
\end{align}
where the ellipsis indicates higher-dimensional operators with more than four derivatives. The functions $g_{IJ}(\phi)$, $V(\phi)$, and $\lambda_{IJKL}(\phi)$ depend on the scalar fields and the couplings in the theory. Note that this is a completely generic parameterization of any scalar EFT as a derivative expansion; we make no assumptions about the symmetries or the dynamics of the theory.  In particular, this expresses the most general EFT interactions up to this derivative order once integration-by-parts identities and field redefinitions have been applied to eliminate redundant operators; see eq.~(2.10) in ref.~\cite{Helset:2020yio}.

The function $g_{IJ}(\phi)$ that appears as the coefficient of the kinetic term can be used to define a geometry on field space. It is a positive-definite two-index tensor that transforms covariantly under non-derivative field redefinitions. As such, it can be interpreted as a metric on a Riemannian manifold for field space~\cite{Volkov:1973vd}.
Starting from the metric, we can derive various descendant quantities. The Christoffel symbol is defined as
\begin{align}
    \Gamma^{I}_{JK} = \frac{1}{2} g^{IL} \left[ g_{LK,J} + g_{JL,K} - g_{JK,L} \right] \,,
\end{align}
where the inverse metric is obtained through $g^{IJ}g_{JK} = \delta^{I}_{K}$, and we use the standard notation
\begin{align}
g_{IJ,K} = \partial_{K} g_{IJ} = \frac{\partial}{\partial \phi^{K}} g_{IJ}\,.
\end{align}
Using the Christoffel symbol, we can assemble the covariant derivative. For example, the covariant derivative of a vector $\eta^{I}$ is
\begin{align}
\big(\eta^{I}\big)_{;J} = \nabla_{J} \eta^{I} = \Big(\delta^{I}_{ K}\partial_J + \Gamma^{I}_{JK} \Big)\eta^K\,,
\end{align}
where again we use the standard notation. Lastly, the Riemann curvature tensor is 
\begin{align}
    R^{I}_{\;\; JKL} = \left[ \Gamma^{I}_{LJ,K} + \Gamma^{I}_{KM}\Gamma^{M}_{JL} - (K \leftrightarrow L) \right] \,.
\end{align}
These geometric quantities are useful because we can express the scattering amplitudes in terms of them.  It is well known that the LSZ reduction formula for scattering amplitudes is proportional to wavefunction factors that account for the fact that the quantum fields (labeled with uppercase index $I$) are interpolating fields for the physical external particle states (labeled with lowercase index $i$).  In the geometric language, this difference is accounted for by the tetrads $e_{Ii}$, which are defined by 
\begin{align}
g_{IJ}(v) = e_{Ii}(v)e_{J}^{i}(v)\,,
\end{align}
where $v^{I}$ is the vacuum expectation value (VEV) of the scalars, see ref.~\cite{Cheung:2021yog} for details.  The transformation properties of the stripped amplitudes (with the tetrad factors removed) under field redefinitions have also been an area of active recent study, leading to the idea of \textit{on-shell covariance}~\cite{Cohen:2022uuw, Cohen:2023ekv, Cohen:2024bml}. The interpretation of the tetrad as the standard wavefunction factor is a straightforward consequence of the interpolating-field condition,
\begin{align}
\langle \Omega| \phi^I(x) |p, i \rangle = e^I_i(v) e^{-i p\cdot x},
\end{align}
where $|\Omega\rangle$ is the interacting vacuum and $|p,i\rangle$ is the physical one-particle state with momentum $p$ and flavor $i$.  Note that this condition involves the on-shell physical state and must be evaluated at the physical vacuum $v$, which is one way to understand the origin of the on-shell covariance of the stripped amplitudes.

As we will review in the next section, the scalar soft theorem can be stated succinctly using field-space geometry, with the additional benefit of providing a nice physical interpretation of the scalar soft theorem.


\subsection{Geometric soft theorem at tree level}

The tree-level geometric soft theorem for scalar EFTs was derived in ref.~\cite{Cheung:2021yog}, and extended to EFTs with scalars, fermions, and gauge bosons in ref.~\cite{Derda:2024jvo}. The geometry of field space plays a central role in the expression of this relation among amplitudes. Consider the soft limit of the momentum $q$ carried by the massless scalar particle with flavor index $i$. For a general EFT Lagrangian of the form \cref{eq:geoNLSM}, the tree-level soft theorem is
\begin{align}\label{eq:TreeSoftTheorem}
    \lim_{q \rightarrow 0} \cA_{n+1,i_1 \cdots i_n i}^{(0)} = \nabla_{i}\s \cA^{(0)}_{n,i_1 \cdots i_n} + \sum_{a=1}^{n} \frac{\nabla_{i} V_{i_a}^{\;\; j_a}}{(p_a+q)^{2} - m_{j_a}^{2}} e^{q\cdot\partial_{p_a}} \cA^{(0)}_{n,i_1\cdots j_a \cdots i_n} + \mathcal{O}(q)\,,
\end{align}
where $\cA_{n}^{(0)}$ is a tree-level $n$-particle scattering amplitude and $V_{ij} \equiv \nabla_{i}\nabla_{j}V$. For clarity, we note that the comma in the subscript of the amplitudes in this formula separates the label denoting the number of external legs from the list of flavor indices for the external legs. On the right-hand side of \cref{eq:TreeSoftTheorem}, the covariant derivative acts on all couplings and masses in the amplitude, which are viewed as functions of the VEV. Thus, the soft limit probes the neighborhood of the theory in field space.  

Note that the soft theorem takes a particularly simple form for theories without a potential:
\begin{align}
\lim_{q \rightarrow 0} \cA_{n+1,i_1 \cdots i_n i}^{(0)} = \nabla_{i}\s \cA^{(0)}_{n,i_1 \cdots i_n} 
\qquad \text{when }\quad V = 0\,.
\label{eq:TreeSoftTheoremVeq0}
\end{align}
The more intricate structure of the second term that appears in \cref{eq:TreeSoftTheorem} is a consequence of the fact that the amplitudes for theories with potential interactions can have singularities in the $q\to 0$ limit due to propagators that scale like $1/(p\cdot q)$ with no compensating momentum dependence in the numerator.  These are tree-level IR divergences, which do not appear in theories with $V=0$ since the derivatives in the interactions regulate any possible tree-level soft singularities.  The relation between theories that possibly have IR-divergent amplitudes and the structure of the scalar soft theorem is central to the one-loop results presented in the following.  In fact, we will show that \cref{eq:TreeSoftTheoremVeq0} holds up to one-loop order without any modification, while the theories with potential interactions have novel complications that lead to a one-loop-corrected form of \cref{eq:TreeSoftTheorem}.

Finally, we note that the extension of the soft theorem to theories with fermions and gauge bosons in the simplest case with no potential simply amounts to the replacement $\nabla \rightarrow \bar \nabla$, where $\bar \nabla$ is the covariant derivative for the full field space of the theory \cite{Derda:2024jvo}. In this work, we will solely focus on theories with scalars and leave the extension to theories with fermions and gauge bosons for future work.

\section{Soft limit of IR-divergent loops}\label{sec:SoftLimit}
As was already seen at tree level, the form of the scalar soft theorem is sensitive to the presence of IR divergences that appear in the $q\to 0$ limit. In four dimensions, it is well known that the $S$-matrix for theories with massless particles can diverge in the IR. We can also encounter UV divergences, which are accommodated following textbook methods of renormalization by invoking local counterterms.  The UV divergences and associated counterterms will not play a role in the arguments presented here.  On the other hand, keeping track of IR divergences is critically important to understand the scalar soft theorem. 

To regulate both UV and IR divergences, we will use dimensional regularization with $d=4-2\epsilon$. The resulting regulated scattering amplitudes are then expanded around $\epsilon=0$, and the expressions generically contain poles in $\epsilon$.  As emphasized above, we must keep track of IR divergences.  The new feature beyond tree-level are IR-divergent loop integrals. Of course, one way to avoid all these complications is to consider theories whose amplitudes are all IR finite. For example, if the loops involve massive particles, then the IR divergences are regulated by the mass.  However, considering a scalar theory without IR divergences is not sufficient. The scalar soft theorem probes a neighborhood of the theory in field space, and the entire neighborhood must be free of IR divergences for the tree-level soft theorem to survive at loop level.  We will argue below that derivatively-coupled theories (and their neighborhoods) fall into this class of IR-finite theories.

More concretely, the key issue is that when we have IR-divergent loops, it is not guaranteed that taking the momenta of one or more massless legs to zero commutes with the evaluation of the loop integral~\cite{Bern:1995ix}. This is intuitive, since both the loop integration and the soft limit probe IR regions.
As we will show explicitly, there are \emph{discontinuities} that can appear when comparing integrated results before and after taking the soft limit.

\begin{figure}[t!]
    \centering
    \begin{subfigure}{0.25\textwidth}
    \begin{tikzpicture}[baseline={([yshift=-2.0ex]current bounding box.center)},scale=0.8]
        \coordinate (a) at (-0.7,0);
        \coordinate (c) at (0.7,0);
        \coordinate (d) at (0,2);
        \tkzDefEquilateral(a,c)\tkzGetPoint{b};
  
        \coordinate (i1) at (-1.5,-0.5);
        \coordinate (i2) at (-1.5,0.5);
        \coordinate (i3) at (-0.5,2);
        \coordinate (i4) at (0.5,2);
        \coordinate (i5) at (1.5,0.5);
        \coordinate (i6) at (1.5,-0.5);

        \draw [hard] (a) -- (b) -- (c) -- (a);
        \draw [hard] (a) -- (i1);
        \draw [hard] (a) -- (i2);
        \draw [hard] (b) -- (i3);
        \draw [hard] (b) -- (i4);
        \draw [hard] (c) -- (i5);
        \draw [hard] (c) -- (i6);
    \end{tikzpicture} 
    \caption{Three-mass\\ triangle integral}
    \label{fig:ThreeMassTriangle}
    \end{subfigure}
    \qquad
    \begin{subfigure}{0.25\textwidth}
    \begin{tikzpicture}[baseline={([yshift=-2.0ex]current bounding box.center)},scale=0.8]
        \coordinate (a) at (-0.7,0);
        \coordinate (c) at (0.7,0);
        \coordinate (d) at (0,2);
        \tkzDefEquilateral(a,c)\tkzGetPoint{b};
  
        \coordinate (i1) at (-1.5,-0.5);
        \coordinate (i2) at (-1.5,0.5);
        \coordinate (i3) at (-0.5,2);
        \coordinate (i4) at (0.5,2);
        \coordinate (i5) at (1.5,0.5);
        \coordinate (i6) at (1.5,-0.5);

        \draw [hard] (a) -- (b) -- (c) -- (a);
        \draw [hard] (a) -- (i1);
        \draw [hard] (b) -- (i3);
        \draw [hard] (b) -- (i4);
        \draw [hard] (c) -- (i5);
        \draw [hard] (c) -- (i6);
     \end{tikzpicture} 
    \caption{Two-mass\\ triangle integral}
    \label{fig:TwoMassTriangle}
    \end{subfigure}
    \qquad
    \begin{subfigure}{0.25\textwidth}
    \begin{tikzpicture}[baseline={([yshift=-2.0ex]current bounding box.center)},scale=0.8]
        \coordinate (a) at (-0.7,0);
        \coordinate (c) at (0.7,0);
        \coordinate (d) at (0,2);
        \tkzDefEquilateral(a,c)\tkzGetPoint{b};
  
        \coordinate (i1) at (-1.5,-0.5);
        \coordinate (i2) at (-1.5,0.5);
        \coordinate (i3) at (-0.5,2);
        \coordinate (i4) at (0.5,2);
        \coordinate (i5) at (1.5,0.5);
        \coordinate (i6) at (1.5,-0.5);

        \draw [hard] (a) -- (b) -- (c) -- (a);
        \draw [hard] (a) -- (i1);
        \draw [hard] (b) -- (i3);
        \draw [hard] (b) -- (i4);
        \draw [hard] (c) -- (i6);
    \end{tikzpicture} 
    \caption{One-mass\\ triangle integral}
    \label{fig:OneMassTriangle}
    \end{subfigure}
    \caption{One-loop triangle integrals of massless scalars, which are functions of (a) one, (b) two, or (c) three external massive momenta. The massless limits (which is the same as taking the soft limit for one of the external legs) of these integrals produce discontinuities when comparing the limit before and after performing the integration.}
    \label{fig:OneLoopTriangleIntegrals}
\end{figure}
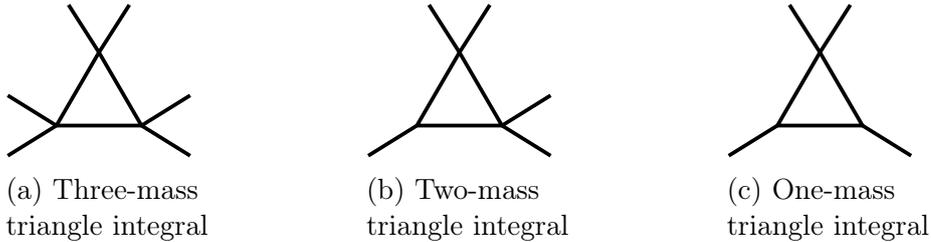

As an example of this phenomenon, consider the one-loop triangle integral with massive external legs and massless internal propagators, $I^{3m}_{3}$, shown in \cref{fig:ThreeMassTriangle}. This integral shows up, for example, in the six-particle one-loop scattering amplitude for a theory with quartic interactions such as $\phi^{4}$-theory.\footnote{Studying loop integrals generated by quartic interactions will be useful for our discussion of the one-loop soft theorem in \cref{sec:OneLoopSoftTheorem}. The case of cubic interactions is postponed to \cref{sec:1loopPotential}.} The three-mass triangle integral is finite. Now, consider taking the limit where one of the external momenta becomes massless. This corresponds to taking the soft limit for one of the external particles for the diagram in \cref{fig:ThreeMassTriangle}. Based on the tree-level result in \cref{eq:TreeSoftTheorem}, one might guess that this limit of the integral is equal to the two-mass triangle integral, $I^{2m}_{3}$, shown in \cref{fig:TwoMassTriangle} and given explicitly in \cref{app:Integrals}. However, this cannot be the case since the two-mass triangle integral contains IR poles in $\epsilon$, while the three-mass triangle integral is finite as $\epsilon \to 0$. Explicitly, taking the limit $k_3^2 \rightarrow 0$ of the integrated results yields \cite{Bern:1995ix}
\begin{align}
    I^{3m}_{3}(k_1^2,k_2^2,k_3^2) \xrightarrow{k^2_3 \to 0} I^{2m}_{3}(k_1^2,k_2^2) - d_2(k_3^2; k_1^2, k_2^2) + d_2(k_3^2; k_2^2, k_1^2)\,,
\end{align}
where the discontinuity $d_{2}$ is given by
\begin{align}
    d_{2}(k_3^2; k_1^2, k_2^2) =\frac{-i}{(4\pi)^{2-\epsilon}}\bigg(\frac{(-k_3^2)^{-\epsilon}}{2\epsilon^2}  - \frac{(-k^2_3)^{-\epsilon}(-k_1^2)^{-\epsilon}}{2\epsilon^2(-k_2^2)^{-\epsilon}} - \textrm{Li}_{2}\left( 1 - \frac{k_1^2}{k_2^2}\right) + \mathcal{O}(\epsilon)  \bigg)\,. \nn
\end{align}
Note that this discontinuity function diverges as a power of $\log k_3^2$ when we first send $\epsilon \to 0$.  This fact will be critical to our arguments below.

Similarly, the massless limit $k_2^2 \rightarrow 0$ of the two-mass triangle integral is discontinuous: 
\begin{align}
    I^{2m}_{3}(k_1^2,k_2^2) \xrightarrow{k^2_2 \to 0} I^{1m}_{3}(k_1^2) - d_1(k_2^2) \,,
\end{align}
where $I^{1m}_{3}$ is the one-mass triangle integral in \cref{fig:OneMassTriangle}, and the discontinuity in this case is given by
\begin{align}
    d_{1}(k^2) = \frac{-i}{(4\pi)^{2-\epsilon}}\frac{1}{\epsilon^2}(-k^2)^{-\epsilon} \,.
\end{align}
The explicit evaluations of the triangle integrals are given in \cref{app:Integrals}. Again, we see that this diverges as a power of $\log k^2$.
These examples illustrate that a soft theorem should potentially be corrected to accommodate such discontinuities in the loop integrals.

Crucially, the discontinuities between the massive and massless one-loop integral can be captured by universal discontinuity functions \cite{Bern:1995ix}, where $d_{1}$ and $d_{2}$ are two examples. These discontinuities in the loop integrals are linked to IR divergences. They are generic features of quantum field theory, and modify the soft theorem in theories with massless particles, such as QCD \cite{Catani:2000pi}.  As we have already emphasized, these functions scale logarithmically with $k^2$ in the limit $\epsilon \to 0$, a property that will be important below. All other one-loop discontinuity functions, such as the massive box integral, also diverge at most logarithmically as one of the external momenta is taken massless. This scaling can be obtained explicitly from the relation between the discontinuity of the box and triangle integrals \cite{Bern:1992em}.
Another potential source of kinematic poles comes from the reduction of a general one-loop integral to the basis of box, triangle, bubble, and tadpole scalar integrals. The coefficient arising from the reduction step may contain dependence on the kinematic variable we want to send to zero. However, in a theory with quartic and higher-point vertices, there are no such poles in the coefficients of the scalar integrals~\cite{Bern:1995ix}. Since the integrals scale logarithmically with the kinematic variable and there are no additional sources of poles as these variables are sent to zero, we can conclude that any one-loop integral arising from a theory with quartic interactions at worst diverges logarithmically in the soft limit. We will make use of this fact to explore the one-loop scalar soft theorem below.

\section{Theories with derivative interactions}
\label{sec:OneLoopSoftTheorem}
In the previous section, we have seen that taking the soft limit of IR-divergent loop integrals does not always commute with performing the integration. Here, we study these subtleties in the context of concrete models, with the goal of extending the geometric soft theorem to one-loop order. This is consistent with the all-orders argument given in sec.~3.4 in the published version of ref.~\cite{Cheung:2021yog}. As anticipated, we will first study the problem with only derivative interactions, and later turn to theories with a nonzero potential. 


The archetypical example of a scalar soft theorem is the Adler zero \cite{Adler:1964um}, which applies to theories of Goldstone bosons with a symmetric coset. The Adler zero states that the soft limit of the scattering amplitude vanishes. This result is valid to all-loop orders \cite{Adler:1964um,Charap:1970xj}. (Recent work has extended this to the loop integrand \cite{Bartsch:2024ofb}.) This feature indicates that the theory of Goldstone bosons does not suffer from IR divergences. It is natural to expect that even theories of Goldstone bosons with nonsymmetric cosets and theories with higher-derivative interactions are free from IR divergences. Thus, as long as we can set the potential to zero, $V(\phi)=0$, the tree-level soft theorem should remain valid to all-loop orders.

The first step in this direction will be taken in this section; we will prove that the geometric soft theorem remains unchanged at one loop for derivatively-coupled theories of the type defined in \cref{eq:geoNLSM}:
\begin{align}\label{eq:1loopst}
    \lim_{q\rightarrow0} \cA^{(1)}_{n+1,i_1 \cdots i_n i} = \nabla_{i} \cA^{(1)}_{n,i_1 \cdots i_n} \,,
\end{align}
where $\cA^{(1)}_{n}$ is a one-loop $n$-particle scattering amplitude, and the soft leg has flavor $i$. We use dimensional regularization to regulate divergences,
\begin{align}
    \cA^{(1)}_{n} = \int\frac{\text{d}^{4-2\epsilon}\ell}{(2\pi)^{4-2\epsilon}} \intA^{(1)}_{n} \,,  
\end{align}
where $\intA^{(1)}_{n}$ is the one-loop integrand. We have absorbed the renormalization scale into the loop integrand for simplicity.

We can more efficiently analyze the soft limit of the one-loop amplitudes in this theory by making a convenient choice of field basis. The scattering amplitudes are invariant under field redefinitions \cite{Chisholm:1961tha, Kamefuchi:1961sb, tHooft:1973wag, Coleman:1969sm, Callan:1969sn, Deans:1978wn, Politzer:1980me, Arzt:1993gz, Passarino:2016saj, Criado:2018sdb, Cohen:2024fak}. The scattering amplitudes for theories of the form \cref{eq:geoNLSM} can be expressed in terms of tensors in field space such as the Riemann curvature tensor and covariant derivatives thereof~\cite{Volkov:1973vd}. However, the loop integrand does not share this property. Typically, the loop integrand is built from a combination of a tensorial part and a non-tensorial part; the non-tensorial terms vanish upon performing the loop integration. To sidestep this complication, we choose to work in \emph{Riemann normal coordinates}, where the full loop integrand is built solely from tensorial terms \cite{Volkov:1973vd}. With this choice, it is easier to make sense of expressions such as $\nabla \intA_{n}^{(1)}$, since every term in the integrand is a tensor.  As a bonus, there are no cubic interactions in this field basis, which makes the analysis significantly simpler. 

Having chosen the field basis that corresponds to Riemann normal coordinates, we now proceed to prove the soft theorem for derivatively-coupled theories of the form \cref{eq:geoNLSM} at one-loop order. The proof is structured in three steps: 

\clearpage
\begin{subequations}
\begin{enumerate}
    \item The geometric soft theorem is valid at the level of the integrand: 
    \begin{equation}\label{eq:Point1}
        \int\frac{\text{d}^{4-2\epsilon}\ell}{(2\pi)^{4-2\epsilon}}\lim_{q\rightarrow0} \intA^{(1)}_{n+1,i_1 \cdots i_n i} = \int\frac{\text{d}^{4-2\epsilon}\ell}{(2\pi)^{4-2\epsilon}} \nabla_{i} \intA^{(1)}_{n,i_1 \cdots i_n} \,.
    \end{equation}
    \item Acting with the covariant derivative commutes with the integration:
    \begin{equation}\label{eq:Point2}
          \int \frac{\text{d}^{4-2\epsilon}\ell}{(2\pi)^{4-2\epsilon}} \nabla_{i} \intA^{(1)}_{n,i_1 \cdots i_n} = \nabla_{i}\int \frac{\text{d}^{4-2\epsilon}\ell}{(2\pi)^{4-2\epsilon}}  \intA^{(1)}_{n,i_1 \cdots i_n}\,.
    \end{equation}
    \item The soft limit commutes with the integration:
    \begin{equation}\label{eq:Point3}
          \int \frac{\text{d}^{4-2\epsilon}\ell}{(2\pi)^{4-2\epsilon}} \lim_{q\rightarrow0} \intA^{(1)}_{n+1,i_1 \cdots i_n i} = \lim_{q\rightarrow0}\int \frac{\text{d}^{4-2\epsilon}\ell}{(2\pi)^{4-2\epsilon}}  \intA^{(1)}_{n+1,i_1 \cdots i_n i}\,.
    \end{equation}
\end{enumerate}
\end{subequations}
If these three conditions hold, then \cref{eq:1loopst} follows directly.

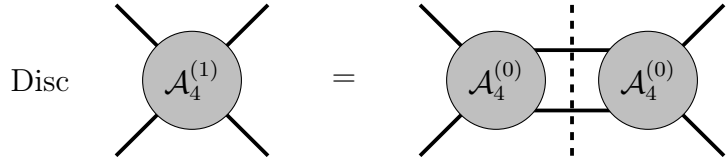
\begin{figure}
    \centering
    \begin{tikzpicture}
            \coordinate (c) at (-3, 1);	
            \coordinate (d) at (-3, -1);	
	    	\coordinate (a) at (-2, 1);
            \coordinate (i) at (-6, 0);
            \coordinate (i1) at (-7, 1);
            \coordinate (i2) at (-7, -1);
            \coordinate (i3) at (-5, 1);
            \coordinate (i4) at (-5, -1);
            \coordinate (j) at (-4, 0);
            \coordinate (j2) at (-8, 0);
            \coordinate (a) at (-1, 1);
            \coordinate (b) at (-1, -1);	
            
	    	
	    	\coordinate (f) at (1, 1);
	    	\coordinate (l) at (1, -1);
	    	
	    	\coordinate (m4) at (-2,0.4);
            \coordinate (m3) at (-2,-0.4);
            \coordinate (m2) at (-2,0);
	    	
	    	\coordinate (mn) at (0, 0);
            \coordinate (mn1) at (-0.25, 0.4);
            \coordinate (mn2) at (-0.25, -0.4);
	    
	    	\coordinate (d1) at (1.00, 0);
	    	\coordinate (d2) at (0.85, 0.5);
	    	\coordinate (d3) at (0.85, -0.5);
	    	
	    	\draw [hard] (c) -- (m2);
	    	\draw [hard] (mn1) -- (m4);

            \draw [hard] (d) -- (m2);
	    	\draw [hard] (mn2) -- (m3);
	    	
	    	\draw [hard] (f) -- (mn);
	        \draw [hard] (l) -- (mn);

            \draw [hard] (i1) -- (i);
	        \draw [hard] (i2) -- (i);
            \draw [hard] (i3) -- (i);
	        \draw [hard] (i4) -- (i);
            
            \draw [hard] (m4) -- (m3);
            \draw [soft] (a) -- (b);
	    	
            \draw[fill=lightgray, opacity=1] (mn) circle (0.65);
            \draw[fill=lightgray, opacity=1] (m2) circle (0.65);
            \draw[fill=lightgray, opacity=1] (i) circle (0.65);
            \node at (m2) {$\cA^{(0)}_{4}$};
            \node at (mn) {$\cA^{(0)}_{4}$};
            \node at (i) {$\cA^{(1)}_{4}$};
            \node at (j) {$=$};
            \node at (j2) {Disc};
                	
\end{tikzpicture}
    \caption{The discontinuity of the one-loop amplitude across a branch cut is given by a product of two tree-level amplitudes. This relation can be leveraged to extract the cut-constructible part of the one-loop integrand using the method of generalized unitarity.}
    \label{fig:Unitarity}
\end{figure}

To show \cref{eq:Point1}, we will recycle on-shell data of tree-level scattering amplitudes to construct the one-loop amplitude.
The one-loop integrand for a general $n$-particle scattering amplitude can be related to products of tree-level scattering amplitudes using the generalized unitarity method \cite{Bern:1994zx,Bern:1994cg}; see \cref{fig:Unitarity} for a four-particle example. The geometric soft theorem holds for tree-level scattering amplitudes. Therefore, combining the tree-level soft theorem with generalized unitarity, we see that the soft theorem automatically holds at the level of the loop integrand, up to terms that vanish upon integration.\footnote{This is consistent with the result in ref.~\cite{Cachazo:2014dia}, which finds that soft theorems in gravity do not receive loop corrections if one takes the soft limit of the loop integrand prior to integration. We, however, are interested in the opposite order of limits, which is why we must also show that \cref{eq:Point2,eq:Point3} hold.} This establishes \cref{eq:Point1}. 

Next, we need to show that the covariant derivative commutes with the integration. We can swap the two operations, shown in \cref{eq:Point2}, because the two operations are independent for theories without potential interactions. The covariant derivative acts solely on the VEVs in the tensors, which are kinematic-independent coefficients of the loop integrand. Conversely, the loop integration depends on the kinematic function of the integrand and not on the coefficient. Therefore, we are free to change the order of operations.\footnote{Note that the situation is different in the case where the theory includes potential interactions or massive particles, since acting with the covariant derivative on a massive propagator would change the analytic function of the integrand. The discussion of these complications are postponed to the section on potential interactions.}

To show \cref{eq:Point3} requires a detailed analysis. For simplicity, let us consider a generic two-derivative theory. The result will straightforwardly generalize to all higher-derivative interactions, since more derivatives make the interactions vanish even more quickly in the soft limit. Recall that we have chosen a field basis that corresponds to Riemann normal coordinates, so the lowest-point vertex has 4 legs.

The soft particle is attached to some vertex in the one-loop amplitude. The line corresponding to the soft particle falls into one of three categories:
\begin{itemize}
    \item[\textbf{a)}] The soft leg is not directly attached to the loop.
    \item[\textbf{b)}] The soft leg is attached to the loop, via an $n$-point vertex where $n\geq5$.
    \item[\textbf{c)}] The soft leg is attached to the loop, via a 4-point vertex.
\end{itemize}
%
%
\begin{figure}[t!]
    \centering
    \begin{subfigure}{0.25\textwidth}
    \begin{tikzpicture}[baseline={([yshift=-2.0ex]current bounding box.center)},scale=0.8]
        \coordinate (a) at (-0.7,0);
        \coordinate (c) at (0.7,0);
        \coordinate (d) at (0,2);
        \tkzDefEquilateral(a,c)\tkzGetPoint{b};
  
        \coordinate (i1) at (-1.5,-0.5);
        \coordinate (i2) at (-1.5,0.5);
        \coordinate (i3) at (-0.5,2);
        \coordinate (i4) at (0.5,2);
        \coordinate (i5) at (1.5,0.5);
        \coordinate (i6) at (1.5,-0.5);
        \coordinate (i7) at (1.5,0.);
        \coordinate (i8) at (1,2.5);
        \coordinate (i9) at (0.5,2.5);
        \coordinate (i10) at (1,2);

        \draw [hard] (a) -- (b) -- (c) -- (a);
        \draw [hard] (a) -- (i1);
        \draw [hard] (a) -- (i2);
        \draw [hard] (b) -- (i3);
        \draw [hard] (b) -- (i4);
        \draw [hard] (c) -- (i5);
        \draw [hard] (c) -- (i6);
        \draw [hard] (c) -- (i7);
        \draw [soft] (i4) -- (i8);
        \draw [hard] (i4) -- (i9);
        \draw [hard] (i4) -- (i10);
    \end{tikzpicture} 
    \caption{}
    \label{fig:SoftInsertionA}
    \end{subfigure}
    \qquad
    \begin{subfigure}{0.25\textwidth}
    \begin{tikzpicture}[baseline={([yshift=-2.0ex]current bounding box.center)},scale=0.8]
        \coordinate (a) at (-0.7,0);
        \coordinate (c) at (0.7,0);
        \coordinate (d) at (0,2);
        \tkzDefEquilateral(a,c)\tkzGetPoint{b};
  
        \coordinate (i1) at (-1.5,-0.5);
        \coordinate (i2) at (-1.5,0.5);
        \coordinate (i3) at (-0.5,2);
        \coordinate (i4) at (0.5,2);
        \coordinate (i5) at (1.5,0.5);
        \coordinate (i6) at (1.5,-0.5);
        \coordinate (i7) at (1.5,0.);
        \coordinate (i8) at (1,2.5);
        \coordinate (i9) at (0.5,2.5);
        \coordinate (i10) at (1,2);

        \draw [hard] (a) -- (b) -- (c) -- (a);
        \draw [hard] (a) -- (i1);
        \draw [hard] (a) -- (i2);
        \draw [hard] (b) -- (i3);
        \draw [hard] (b) -- (i4);
        \draw [hard] (c) -- (i5);
        \draw [soft] (c) -- (i6);
        \draw [hard] (c) -- (i7);
        \draw [hard] (i4) -- (i8);
        \draw [hard] (i4) -- (i9);
        \draw [hard] (i4) -- (i10);
    \end{tikzpicture} 
    \caption{}
    \label{fig:SoftInsertionB}
    \end{subfigure}
    \qquad
    \begin{subfigure}{0.25\textwidth}
    \begin{tikzpicture}[baseline={([yshift=-2.0ex]current bounding box.center)},scale=0.8]
        \coordinate (a) at (-0.7,0);
        \coordinate (c) at (0.7,0);
        \coordinate (d) at (0,2);
        \tkzDefEquilateral(a,c)\tkzGetPoint{b};
  
        \coordinate (i1) at (-1.5,-0.5);
        \coordinate (i2) at (-1.5,0.5);
        \coordinate (i3) at (-0.5,2);
        \coordinate (i4) at (0.5,2);
        \coordinate (i5) at (1.5,0.5);
        \coordinate (i6) at (1.5,-0.5);
        \coordinate (i7) at (1.5,0.);
        \coordinate (i8) at (1,2.5);
        \coordinate (i9) at (0.5,2.5);
        \coordinate (i10) at (1,2);

        \draw [hard] (a) -- (b) -- (c) -- (a);
        \draw [soft] (a) -- (i1);
        \draw [hard] (a) -- (i2);
        \draw [hard] (b) -- (i3);
        \draw [hard] (b) -- (i4);
        \draw [hard] (c) -- (i5);
        \draw [hard] (c) -- (i6);
        \draw [hard] (c) -- (i7);
        \draw [hard] (i4) -- (i8);
        \draw [hard] (i4) -- (i9);
        \draw [hard] (i4) -- (i10);
    \end{tikzpicture} 
    \caption{}
    \label{fig:SoftInsertionC}
    \end{subfigure}
    \caption{Examples of one-loop diagrams with all possible insertions of the soft particle, denoted with a dashed line. a) The soft particle vertex is not directly part of the loop. b) The soft particle is attached to the loop, but with a higher-point vertex. c) The soft particle is attached to the loop with a 4-point vertex.}
    \label{fig:SoftInsertions}
\end{figure}
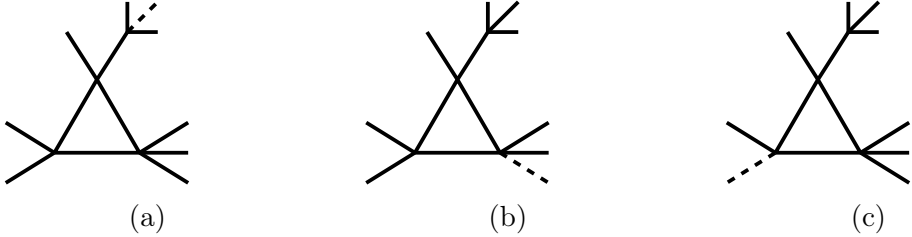

These three possibilities are shown in \cref{fig:SoftInsertions}. Now, consider the commutation of the soft limit and loop integration. For case \textbf{a}, clearly the two limits commute since the soft vertex is displaced away from the loop itself. Case \textbf{b} also poses no issues with the order of limits.  The reason is that the vertex with the soft leg attached has three or more lines that are external to the loop, and so the total momentum flowing into the vertex remains massive after taking the soft limit.  Therefore, the analytic structure of the loop integrals is unchanged even after taking the soft limit. For example, if the vertex was part of a three-mass triangle integral, then it would remain a three-mass triangle integral after taking the soft limit. Thus, loop integrals of this type have a smooth soft limit.

The last option, case \textbf{c}, is that the soft particle is attached to a four-point vertex in the loop. We have already seen in \cref{sec:SoftLimit} that the soft limit of integrals built from four-point vertices may lead to discontinuities. We additionally reviewed that these discontinuities scale at most logarithmically with the soft momentum.
However, this possible complication is avoided here, since the numerator will always be proportional to the soft momentum. We can see why this is the case by looking at the explicit form of the four-point Feynman rule in normal coordinates: 
\begin{equation}
   \hspace{-2pt} 
   \begin{tikzpicture}[baseline={([yshift=-1.0ex]current bounding box.center)},scale=0.8]
        \coordinate (a) at (0,0);
        
        \coordinate (perm) at (3,0.5);
        \coordinate (i1) at (-1,-1);
        \coordinate (i2) at (-1,1);
        \coordinate (i3) at (1,1);
        \coordinate (i4) at (1,-1);

        \node[left]  at (i1) {$i$};
        \node[left]  at (i2) {$i_a$};
        \node[right]  at (i3) {$i_b$};
        \node[right]  at (i4) {$i_c$};

        \draw [soft] (a) -- (i1);
        \draw [hard] (a) -- (i2);
        \draw [hard] (a) -- (i3);
        \draw [hard] (a) -- (i4);
    \end{tikzpicture} 
  = i \bigg[ R_{ii_ai_bi_c} \big(2\s q \cdot p_b +\frac{1}{3}(2\s p_b^2-p_a^2-p_c^2)\big) + R_{ii_bi_ai_c}\big( 2\s q \cdot p_a +\frac{1}{3}(2\s p_a^2-p_b^2-p_c^2)\big)  \bigg]\,,
\end{equation}
where the soft momentum $q$ is associated with the particle with flavor $i$ as above.
From this expression, we see that the vertex is either proportional to the soft momentum $q$, which implies that the diagram vanishes in the soft limit, or it is proportional to the square of a momentum, which either vanishes on-shell or cancels a propagator that is attached to the vertex.  This second case reduces to a diagram where the soft particle is attached to a higher-point vertex, so it is covered by case \textbf{b}. Thus, the soft limit of such diagrams commutes with the loop integration, thereby demonstrating \cref{eq:Point3}. 

The same logic applies to any higher-derivative interaction. The crucial properties we used for the two-derivative theory are that: 1) we can use a field basis with no cubic interactions, and 2) the on-shell quartic vertex scales with the soft momentum in the soft limit. For any higher-derivative interaction, both properties can easily be made manifest. We have therefore shown that the geometric soft theorem remains valid for all derivatively-coupled theories at one-loop order. These results have been verified in numerous examples, some of which are given in \cref{app:Goldstones}.

\section{Theories with potential interactions}\label{sec:1loopPotential}
Now we turn to analyze theories that include potential interactions which do not involve derivatives. 
We revisit the arguments in the previous subsection to see how the conclusions change in the presence of potential interactions.  For simplicity, we assume all the particles in the theory are massless in this section.  The main result in this section will be to derive the form of the leading one-loop correction to \cref{eq:TreeSoftTheorem}.

First, note that the soft theorem in \cref{eq:TreeSoftTheorem} remains valid for the one-loop integrand.  This is because we can build the loop integrand from tree-level amplitudes using generalized unitarity, even in the presence of potential interactions. We emphasize that the right-hand side of \cref{eq:Point1} must be modified to include the potential term in \cref{eq:TreeSoftTheorem}.

Next, we must consider the commutation of the soft limit and the loop integration. As before, we analyze all possible insertions of the vertex with the soft particle. The possibilities include the cases \textbf{a}, \textbf{b}, and \textbf{c} from the previous section, along with a new option, namely
\begin{itemize}
    \item[\textbf{d)}] The soft leg is attached to the loop with a 3-point vertex.
\end{itemize}
When considering modifications at one loop, the potential interactions $\phi^{n}$ can be split into three classes: higher-point interactions with $n\geq 5$, quartic interactions with $n=4$, and cubic interactions with $n=3$. We analyze each group in turn.
 
Potential interactions $\phi^n$ with $n\geq 5$ \textit{do not} modify the soft theorem at one-loop order. This is because the arguments above for points \textbf{a} and \textbf{b} go through unchanged. In addition, in the absence of cubic or quartic interactions, points \textbf{c} and \textbf{d} do not apply. Note that in this case, the soft theorem is fully captured by the covariant derivative, as in \cref{eq:1loopst}.

If the theory includes a $\phi^4$ interaction, the situation is different. The above arguments for points \textbf{a} and \textbf{b} are still valid, but we face an issue with point \textbf{c}. The loop integrals built from quartic vertices do not have a smooth soft limit. For derivatively-coupled scalars, this was overcome by the momentum dependence in the numerator of the vertex. However, for quartic potential interactions, there is no compensating kinematic factor in the vertex, and the soft limit of the loop integrals diverge logarithmically. The soft theorem must therefore be modified to account for this class of interactions.

Theories with $\phi^3$ interactions lead to even more dramatic differences. The cubic interactions allow for new IR-divergent one-loop integrals as well as divergences from tree-level propagators.  This implies that the arguments for the commutation of the soft limit and loop integration, \cref{eq:Point3}, completely fail for theories with cubic potential interactions. The leading divergence scale as $1/(p_a\cdot q)^2$ at one-loop order, compared to $1/(p_a\cdot q)$ at tree level. However, since the failure of the tree-level soft theorem is linked to IR divergences, there is a hope that the violation may be universal.  Exploring the extent to which there is a universal correction to the soft theorem is the topic we turn to next.


The soft theorem must be modified when working with theories that include potential interactions.  Here, we derive the leading one-loop correction to the soft theorem for a massless theory with a $\phi^3$ interaction. Our strategy will closely follow the derivation of the analogous result for QCD amplitudes~\cite{Catani:2000pi}. At tree level, the soft limit is governed by the geometric soft theorem in \cref{eq:TreeSoftTheorem}. It constitutes the covariant derivative in field space and the contribution from the tree-level insertion of a cubic interaction on external legs. The second term diverges as $1/(p_a\cdot q)$ in the soft limit.

We will extend this result to one-loop amplitudes, $\cA^{(1)}$. As we shall see, the leading part of the soft limit takes the \emph{factorized} form
\begin{align}\label{eq:K1}
    \lim_{q\rightarrow 0} \cA^{(1)}_{n+1} = K^{(1)}(q) \cA^{(0)}_{n}
    + \mathcal{O}\big(1/(p_a\cdot q)\big) \,,
\end{align}
where the leading one-loop correction to the soft theorem, $K^{(1)}(q)$, scales as $1/(p_a\cdot q)^2$ in the soft limit. This term is sensitive to the noncommutation of the soft limit and the loop integration due to IR divergences.

There are two types of contributions to \cref{eq:K1}; diagrams involving a single hard line, shown in \cref{fig:Bubble,fig:OneHardLine}, and diagrams involving two hard lines, shown in \cref{fig:TwoHardLines}. The former contributions obviously factorize, while to establish the factorization of the latter contributions requires a slightly deeper analysis.  We divide the one-loop correction to the soft theorem as
\begin{align}
    K^{(1)}(q)\cA_{n}^{(0)} = \sum_{a} K^{(1)}_{a}(q) \cA^{(0)}_{n} + \sum_{a,b} K^{(1)}_{a,b}(q) \cA^{(0)}_{n} \,,
\end{align}
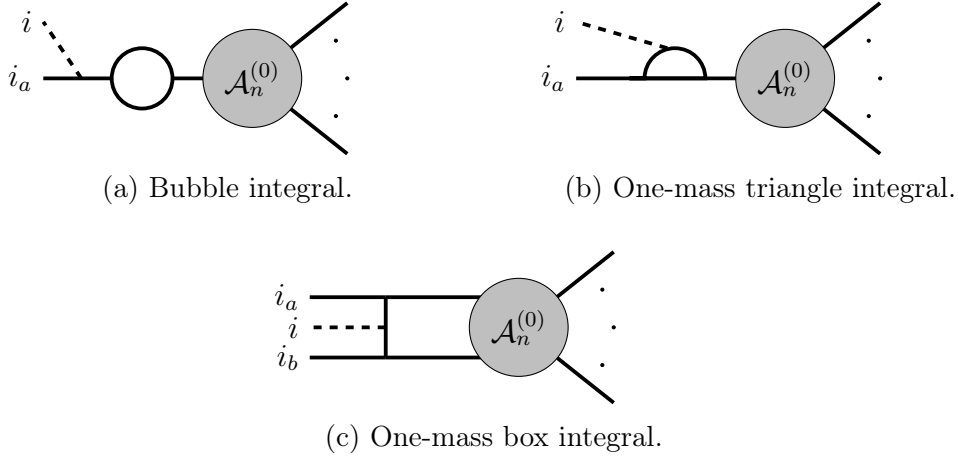
\begin{figure}
    \centering
    \begin{subfigure}{0.4\textwidth}
    \begin{tikzpicture}
            \coordinate (c) at (-3, 0);	
	    	\coordinate (a) at (-2, 1);
            \coordinate (i) at (-3, 0.75);
            
	        \node[left]  at (c) {$i_a$};
            \node[left]  at (i) {$i$};
	    	
	    	\coordinate (f) at (1, 1);
	    	\coordinate (l) at (1, -1);
	    	
	    	\coordinate (m4) at (-2.5,0);
            \coordinate (m5) at (-2.1,0);
            \coordinate (m3) at (-1.3,0);
            \coordinate (m2) at (-1.8,0.4);
	    	
	    	\coordinate (mn) at (-0.25, 0);
	    
	    	\coordinate (d1) at (1.00, 0);
	    	\coordinate (d2) at (0.85, 0.5);
	    	\coordinate (d3) at (0.85, -0.5);
	    	
	    	\draw [hard] (c) -- (m5);
	    	\draw [hard] (mn) -- (m3);
	    	
	    	\draw [hard] (f) -- (mn);
	        \draw [hard] (l) -- (mn);

            \draw [hard] (m3) arc(0:360:0.4) --cycle;
            \draw [soft] (m4) -- (i);
      
            \draw[fill=lightgray, opacity=1] (mn) circle (0.65);
            \node at (mn) {$\cA^{(0)}_{n}$};
                	
            \draw[fill=black, opacity=1] (d1) circle (0.02);
            \draw[fill=black, opacity=1] (d2) circle (0.02);
            \draw[fill=black, opacity=1] (d3) circle (0.02);
\end{tikzpicture}
    \caption{Bubble integral.}
    \label{fig:Bubble}
    \end{subfigure}
    \qquad
    \begin{subfigure}{0.4\textwidth}
    \begin{tikzpicture}
            \coordinate (c) at (-3, 0);	
	    	\coordinate (a) at (-2, 1);
            \coordinate (i) at (-3, 0.75);
            
	        \node[left]  at (c) {$i_a$};
            \node[left]  at (i) {$i$};
	    	
	    	\coordinate (f) at (1, 1);
	    	\coordinate (l) at (1, -1);
	    	
	    	\coordinate (m4) at (-2.3,0);
            \coordinate (m3) at (-1.3,0);
            \coordinate (m2) at (-1.8,0.4);
	    	
	    	\coordinate (mn) at (-0.25, 0);
	    
	    	\coordinate (d1) at (1.00, 0);
	    	\coordinate (d2) at (0.85, 0.5);
	    	\coordinate (d3) at (0.85, -0.5);
	    	
	    	\draw [hard] (c) -- (m4);
	    	\draw [hard] (mn) -- (m4);
	    	
	    	\draw [hard] (f) -- (mn);
	        \draw [hard] (l) -- (mn);

            \draw [hard] (m4) -- (m3) arc(0:180:0.4) --cycle;
            \draw [soft] (m2) -- (i);
      
            \draw[fill=lightgray, opacity=1] (mn) circle (0.65);
            \node at (mn) {$\cA^{(0)}_{n}$};
                	
            \draw[fill=black, opacity=1] (d1) circle (0.02);
            \draw[fill=black, opacity=1] (d2) circle (0.02);
            \draw[fill=black, opacity=1] (d3) circle (0.02);
\end{tikzpicture}
    \caption{One-mass triangle integral.}
    \label{fig:OneHardLine}
    \end{subfigure}
    \\[15pt]
    \begin{subfigure}{0.4\textwidth}
    \begin{tikzpicture}
            \coordinate (c) at (-3, 0.4);	
            \coordinate (d) at (-3, -0.4);	
	    	\coordinate (a) at (-2, 1);
            \coordinate (i) at (-3, 0);
            
	        \node[left]  at (c) {$i_a$};
            \node[left]  at (d) {$i_b$};
            \node[left]  at (i) {$i$};
	    	
	    	\coordinate (f) at (1, 1);
	    	\coordinate (l) at (1, -1);
	    	
	    	\coordinate (m4) at (-2,0.4);
            \coordinate (m3) at (-2,-0.4);
            \coordinate (m2) at (-2,0);
	    	
	    	\coordinate (mn) at (-0.25, 0);
            \coordinate (mn1) at (-0.25, 0.4);
            \coordinate (mn2) at (-0.25, -0.4);
	    
	    	\coordinate (d1) at (1.00, 0);
	    	\coordinate (d2) at (0.85, 0.5);
	    	\coordinate (d3) at (0.85, -0.5);
	    	
	    	\draw [hard] (c) -- (m4);
	    	\draw [hard] (mn1) -- (m4);

            \draw [hard] (d) -- (m3);
	    	\draw [hard] (mn2) -- (m3);
	    	
	    	\draw [hard] (f) -- (mn);
	        \draw [hard] (l) -- (mn);

            \draw [hard] (m4) -- (m3);
            \draw [soft] (m2) -- (i);
	    	
            \draw[fill=lightgray, opacity=1] (mn) circle (0.65);
            \node at (mn) {$\cA^{(0)}_{n}$};
                	
            \draw[fill=black, opacity=1] (d1) circle (0.02);
            \draw[fill=black, opacity=1] (d2) circle (0.02);
            \draw[fill=black, opacity=1] (d3) circle (0.02);
\end{tikzpicture}
    \caption{One-mass box integral.}
    \label{fig:TwoHardLines}
    \end{subfigure}
    \caption{One-loop corrections to the leading soft theorem that involves either one or two hard lines.}
    \label{fig:OneLoopCorrections}
\end{figure}%
where we sum over a single or pairs of hard particles. For a single hard line, we find that the contributions, shown in \cref{fig:Bubble,fig:OneHardLine}, are 
\begin{align}
    \sum_{a} K^{(1)}_{a}(q) \cA^{(0)}_{n}
    &=  \sum_{a} I_{2}((p_a+q)^2) \frac{i V_{i_a i}^{\quad j_1} V_{j_1 }^{\;\; j_2j_3} V_{j_2j_3}^{\quad\; j_a}}{2\left[2(p_a \cdot q)\right]^2} e^{q\cdot \partial_{p_a}} \cA^{(0)}_{n, \cdots j_a \cdots}
    \nonumber \\[4pt]
    &\hspace{12pt}+
    \sum_{a} I^{1m}_{3}((p_a+q)^2) \frac{i V_{i_a j_1}^{\quad j_2} V_{i j_2}^{\quad j_3} V_{j_3}^{\;\; j_1 j_a}}{2(p_a \cdot q)} e^{q\cdot \partial_{p_a}} \cA^{(0)}_{n, \cdots j_a \cdots}\,.
    \label{eq:KaTerm}
\end{align}
where $V_{ijk} = \nabla_{i}\nabla_{j}\nabla_{k}V$. The bubble integral, $I_2$, and the one-mass triangle integral, $I^{1m}_{3}$, are given in \cref{app:Integrals}.

The contributions from pairs of hard particles are 
\begin{align}
    \sum_{a, b} K^{(1)}_{a,b}(q) \cA_{n}^{(0)} = &\sum_{a<b} I^{(1)}_{\Box} i V_{i}^{\;\;\; j_1 j_2} V_{i_a j_1}^{\quad j_a} V_{i_b j_2}^{\quad j_b} \cA_{n,\cdots j_a \cdots j_b \cdots}^{(0)} \,,
    \label{eq:KabTerm}
\end{align}
where 
\begin{align}
    I^{(1)}_{\Box} &= I^{1m}_{4}\big((p_a + q)^2,(p_b + q)^2,(p_a+p_b+q)^2\big) \notag\\[4pt] 
    &=\frac{i}{16\pi^2}\frac{1}{2}\frac{1}{p_a\cdot q \, p_b\cdot q}\bigg[\frac{1}{\epsilon^2}-\frac{1}{\epsilon} \log\bigg(-\frac{p_a\cdot q}{\pi\mu^2e^{-\gamma}}\frac{p_b\cdot q}{s_{ab}}\bigg)\notag\\[4pt]
    &\hspace{12pt}+\frac{1}{2}\log^2\bigg(-\frac{p_a\cdot q}{\pi\mu^2e^{-\gamma}}\frac{p_b\cdot q}{s_{ab}}\bigg)+ \frac{\pi^2}{12}
    +\mathcal{O}(\epsilon)+\mathcal{O}(q)  \bigg]\,.
\end{align}
%
where $s_{ab} = (p_a+p_b)^2$ as usual.
The reason why these contributions factorize can be understood by using the method of regions~\cite{Beneke:1997zp}. The leading term in the soft limit is when the loop momentum is in the soft region and the soft particle is inserted between two (hard) massless particles. This configuration is enhanced because the hard particles are near on shell, and because of this, these contributions factorize.

The \emph{leading} contributions to scalar one-loop scattering amplitudes in the soft limit are captured by \cref{eq:KaTerm,eq:KabTerm}.  
These together provide the one-loop corrections to the leading soft theorem for scalar EFTs. 

\section{Examples}\label{app:Examples}
In this section, we will study two concrete examples that demonstrate the general results for the one-loop soft theorem. We will first discuss theories with only derivative couplings, then consider the impact of adding potential interactions. 
\subsection{Two-derivative interactions}\label{app:Goldstones}

We start with the Lagrangian
\begin{align}
	\mathcal{L} = \frac{1}{2} g_{IJ}(\phi) (\partial_{\mu} \phi^{I}) (\partial^{\mu} \phi^{J}) \,.
\end{align}	
From this Lagrangian, we will calculate various scattering amplitudes at loop level.

The four-point one-loop amplitude is 
\begin{align}\label{eq:amplitude4ptR}
	\cA^{(1)}_{4,i_1 i_2 i_3 i_4} &=  i s_{34}I_2(s_{34})\Big\{ \left[ \left( 2R_{i_1 j_1 i_2 j_2} + R_{i_1 i_2 j_1 j_2}\right) R_{i_3 \;\;\; i_4 }^{\;\;\; j_1 \;\;\; j_2 } + R_{i_1 j_1 i_2 j_2}  R_{i_3 i_4}^{\quad j_1 j_2}  \right] s_{12}  \\
	&\qquad\quad+ \left[ R_{i_1 i_2 j_1 j_2} R_{i_3 i_4}^{\quad j_1 j_2} \right] \left(\frac{d}{2(d-1)} s_{12}  + \frac{1}{(d-1)} s_{13} \right) \Big\} + \rm{cycl}(1,2,3) \nonumber\,, 
\end{align}
where $I_2(p^2)$ is the bubble integral in \cref{eq:IntegralBubble}, and we sum over cyclic permutations.

\clearpage
\noindent The five-point amplitude is
\begin{align}
	\cA^{(1)}_{5,i_1 i_2 i_3 i_4 i_5} &= \frac{is_{34}}{6} I_2(s_{34}) \Big\{\Big[ ( \nabla_{i_2} R_{j_2 i_1 j_1 i_5} + \nabla_{i_5} R_{j_2 j_1 i_2 i_1}) s_{15} + \nabla_{i_1} R_{j_2 i_2 j_1 i_5} s_{25} \nonumber \\ & \qquad\qquad\qquad\qquad
    + \nabla_{i_5} R_{j_2 i_2 j_1 i_1} s_{12}     \Big] 
	\Big[  2 R_{i_3 \;\;\; i_4 \;\;\;}^{\;\;\; j_1 \;\;\; j_2} + R_{i_3 i_4}^{\quad j_1 j_2}  \Big]  \nonumber \\	    
    &+  \Big[   \nabla_{i_1} R_{j_2 j_1 i_2 i_5} (s_{35} + s_{45})  +  \nabla_{i_5} R_{j_2 j_1 i_2 i_1} (s_{13} + s_{14})  \Big] 
	\Big[  - R_{i_3 \;\;\; i_4 \;\;\;}^{\;\;\; j_1 \;\;\; j_2}  \Big]  \nonumber \\
		     &+  \Big[   \nabla_{i_1} R_{j_2 j_1 i_2 i_5}    \Big] 
	\Big[ - R_{i_3 i_4}^{\quad j_1 j_2} \Big] \left(\frac{d}{2(d-1)} (s_{35} + s_{45}) - \frac{1}{(d-1)}  s_{35} \right)  \nonumber \\
	     &+  \Big[   \nabla_{i_5} R_{j_2 j_1 i_2 i_1}     \Big] 
	\Big[ - R_{i_3 i_4}^{\quad j_1 j_2} \Big] \left(\frac{d}{2(d-1)} (s_{13}+s_{14}) - \frac{1}{(d-1)}  s_{13} \right) \Big\}  
    \nonumber \\ &
    + \rm{perm}(1,2,3,4,5) \,,
\end{align}	
where we sum over all permutations.

We can now study the scattering amplitudes in the soft limit. First, we look at the $p_{4}\rightarrow 0$ limit of the four-point amplitude. The bubble integral diverges logarithmically in this limit. However, the overall kinematic factor compensates for this, making the soft limit of the amplitude vanish. This agrees with the geometric soft theorem, since there is no nonvanishing three-point amplitude in this theory.

Next, we turn to the five-point amplitude. Consider the $p_{5}\rightarrow 0$ limit of this amplitude. Again, the bubble integral diverges whenever its argument goes to zero. But from the structure of the scattering amplitude, we see that the bubble integral always comes together with a kinematic factor, such as $s_{34} I_{2}(s_{34})$. This makes this limit of the scattering amplitude finite. The scattering amplitude in the soft limit is
\begin{align}
	\lim_{p_5\rightarrow 0} \cA^{(1)}_{5,i_1 i_2 i_3 i_4 i_5} 
	&=  
	i s_{34} I_2(s_{34}) \Big\{ \left[ (\nabla_{i_5} R_{j_2 i_2 j_1 i_1}) \left( 2 R_{i_3 \;\;\; i_4 \;\;\;}^{\;\;\; j_1 \;\;\; j_2} + R_{i_3 i_4}^{\quad j_1 j_2} \right) \right]  s_{12}   \nonumber \\
	&+ \left[ (\nabla_{i_5} R_{j_2 j_1 i_2 i_1}) R_{i_3 \;\;\; i_4 \;\;\;}^{\;\;\; j_1 \;\;\; j_2} \right] (-s_{13} - s_{14})  \nonumber \\
	&+ \left[ (\nabla_{i_5} R_{j_2 j_1 i_2 i_1}) R_{i_3 i_4}^{\quad j_1 j_2} \right]
	\left(\frac{d}{2(d-1)} (-s_{13}-s_{14}) + \frac{1}{(d-1)} s_{13} \right)\Big\}  \nonumber \\ &+ \rm{cycl}(1,2,3) + \left(1,2  \leftrightarrow 3,4 \right) \,.
\end{align}
This is precisely the geometric soft theorem;
\begin{align}
	\lim_{p_5\rightarrow 0} \cA^{(1)}_{5,i_1 i_2 i_3 i_4 i_5} 
	&=  \nabla_{i_5} \cA^{(1)}_{4,i_1 i_2 i_3 i_4} \,.
\end{align}
We also have verified that the geometric soft theorem holds for the six-particle scattering amplitude. In addition, we computed one-loop scattering amplitude for higher-derivative corrections $\lambda (\partial\phi)^4$ up to five particles. The soft limits of these amplitudes all agree with the geometric soft theorem.

\subsection{Potential interactions}
We now study the effect of potential interactions. In particular, we consider the most relevant deformation---the cubic coupling---in a theory with four-derivative interactions and a flat metric. The latter assumption simplifies the diagrammatic analysis without modifying the discussion of the corrections to the soft theorem. Concretely, we consider the Lagrangian 
\begin{align}
    \Lagr = \frac{1}{2} \delta_{IJ} (\partial_{\mu} \phi^{I})(\partial^{\mu} \phi^{J}) - \frac{1}{3!} V_{IJK} \phi^{I}\phi^{J}\phi^{K} + \lambda_{IJKL}(\phi) (\partial_{\mu} \phi^{I})(\partial^{\mu} \phi^{J}) (\partial_{\nu} \phi^{K}) (\partial^{\nu} \phi^{L}) \,. 
\end{align}
Note that $V_{IJK}$ is a coupling constant (it is not a function of the scalar field).
We will now study the soft theorem by isolating contributions with specific numbers of insertions of the cubic coupling, since the soft theorem must hold for each set of diagrams in this class independently. We introduce the notation $\cA^{(1),V^n}_{5}$ for the part of the one-loop amplitude with $n$ insertions of $V_{IJK}$. 

For zero insertions of the cubic coupling, the situation is similar to the one described previously:
\begin{align}
    \lim_{p_5 \rightarrow 0} \cA^{(1),V^0}_{5}  = \nabla_{i_5} \cA^{(1),V^0}_{4}.
\end{align}

Next, we turn to amplitudes with one insertion of the cubic coupling. For four particles, there are no such terms, $\cA^{(1),V^1}_{4} = 0$. For five particles, the amplitude is
\begin{equation}\label{eq:5pointLambda3}
    \cA^{(1),V^1}_{5} = 
    \begin{tikzpicture}[baseline={([yshift=-2.3ex]current bounding box.center)},scale=0.5]
        \coordinate (a) at (-0,0.5);
        \coordinate (c) at (1,0.5);
        \coordinate (d) at (1.5,1);

        \coordinate (p) at (-2,0.5);
        \coordinate (i1) at (-0.5,0);
        \coordinate (i2) at (-0.5,1);
        \coordinate (i3) at (1.5,1.5);
        \coordinate (i4) at (2,1);
        \coordinate (i5) at (1.5,0);

        \node[left]  at (i1) {$i_2$};
        \node[left]  at (i2) {$i_1$};
        \node[above]  at (i3) {$i_5$};
        \node[right]  at (i4) {$i_4$};
        \node[right]  at (i5) {$i_3$};
        \node[left] at (p) {$\frac{1}{8}$};

        \draw (a) -- (a) ;
        \draw (c) arc(0:360:0.5) --cycle;
        \draw (c) -- (d);
        \draw (a) -- (i1);
        \draw (a) -- (i2);
        \draw (d) -- (i3);
        \draw (d) -- (i4);
        \draw (c) -- (i5);
    \end{tikzpicture}
    \quad
    \begin{tikzpicture}[baseline={([yshift=-2.0ex]current bounding box.center)},scale=0.4]
        \coordinate (a) at (-0.7,0);
        \coordinate (c) at (0.7,0);
        \coordinate (d) at (0,2);
        \tkzDefEquilateral(a,c)\tkzGetPoint{b};

        \coordinate (p) at (-3,0.5);
        \coordinate (perm) at (3,0.5);
        \coordinate (i1) at (-1.5,-0.5);
        \coordinate (i2) at (-1.5,0.5);
        \coordinate (i3) at (0,1.7);
        \coordinate (i4) at (1.5,0.5);
        \coordinate (i5) at (1.5,-0.5);

        \node[left]  at (i1) {$i_2$};
        \node[left]  at (i2) {$i_1$};
        \node[above]  at (i3) {$i_5$};
        \node[right]  at (i4) {$i_4$};
        \node[right]  at (i5) {$i_3$};
        \node[left] at (p) {$+\,\frac{1}{8}$};
        \node[right] at (perm) {$+\,\textrm{perm}(1,2,3,4,5)\,,$};

        \draw (a) -- (b) -- (c) -- (a);
        \draw (a) -- (i1);
        \draw (a) -- (i2);
        \draw (b) -- (i3);
        \draw (c) -- (i4);
        \draw (c) -- (i5);
    \end{tikzpicture} 
\end{equation}
where the quartic vertices are the Feynman vertices from $\lambda(\partial\phi)^4$, and the cubic vertex comes from the potential. Consider the soft limit where $p_5\rightarrow 0$. Since we are summing over all permutations of the external particles in \cref{eq:5pointLambda3}, we have to consider how each term in the sum behaves in the soft limit. Whenever particle 5 is attached to a quartic vertex coming from the four-derivative interaction, this diagram vanishes in the soft limit. This is similar to how \cref{eq:amplitude4ptR} behaves in the soft limit. For the bubble graph in \cref{eq:5pointLambda3}, the only other option is for particle 5 to be connected to the cubic vertex. We recognize this graph as the tree-level modification of an external line by the cubic vertex, which is part of the original tree-level soft theorem. In turn, the bubble graph in \cref{eq:5pointLambda3} corresponds to the pole term in the soft theorem in \cref{eq:TreeSoftTheorem}, i.e., $+\sum \frac{V_{ijk}}{(p+q)^2-m^2} \cA^{(1),V^0}_{4}$.

For the second term, we again have two options for where particle 5 is attached; either to a quartic or a cubic vertex. The graphs where particle 5 is attached to a quartic vertex vanish in the soft limit, since the vertex scales as $\mathcal{O}(p_5\cdot p_a)$ and the loop integral scales as $\mathcal{O}(\log p_5\cdot p_a)$. The only nonvanishing term is when particle 5 is attached to the cubic vertex. This corresponds to the soft limit of a massless leg in a two-mass triangle integral. This integral is given in \cref{app:Integrals}. On the right-hand side of the soft theorem, this corresponds to the covariant derivative acting on an internal propagator of a bubble diagram, effectively doubling it. The bubble integral with additional propagator factors is also given in \cref{app:Integrals}. In particular, the soft theorem holds as long as
\begin{align}
    \lim_{s_{34} \rightarrow s_{12}} I^{2m}_{3}\big(s_{12},s_{34}\big) = J_{2,12}\big(s_{12}\big) \,,
\end{align}
which corresponds to the $p_5\rightarrow 0$ due to the 5-point kinematics. 
By combining these two results, we have shown that the soft theorem in \cref{eq:TreeSoftTheorem} also holds for the parts of the one-loop amplitudes that are linear in the cubic coupling: 
\begin{align}
    \lim_{p_5\rightarrow 0}   \cA^{(1),V^1}_{5} = \nabla_{i_5}  \cA^{(1),V^1}_{4}+\sum_{a=1}^{4} \frac{\nabla_{i_5} V_{i_a}^{\;\; j_a}}{2(p_a\cdot p_5)} e^{p_5\cdot\partial_{p_a}} \cA^{(1),V^0}_{4,i_1\cdots j_a \cdots i_4} \,.
\end{align}

Next, we could analyze the part of the scattering amplitudes with two insertions of cubic interactions. They scale as $1/(p_a\cdot p_5)$ in the soft limit. However, these contributions do not exactly match the tree-level soft theorem. Therefore, there must be a one-loop modification to the soft theorem. Such modifications are subleading compared to the results derived in \cref{sec:1loopPotential}, and beyond the scope of this paper. Instead, we will look at the part of the amplitude that is sensitive to the \emph{leading} correction to the soft theorem, namely three insertions of the cubic coupling. 

\clearpage
The diagrammatic representation of the amplitude with three insertions of the cubic coupling is given by 
\begin{equation*}
     \cA^{(1),V^3}_{5} = \tfrac{1}{24}
\begin{tikzpicture}[baseline={([yshift=-2.3pt]current bounding box.center)},scale=0.5]

    \coordinate (a) at (0,0);
    \coordinate (b) at (1,0);
    \coordinate (c) at (2,0);
    \coordinate (d) at (3,0);

    \coordinate (i1) at (-0.5,1);
    \coordinate (i2) at (-0.5,-1);
    \coordinate (i3) at (3.5,-1);
    \coordinate (i4) at (4,0);
    \coordinate (i5) at (3.5,1);

    \node[left]  at (i1) {$i_1$};
    \node[left]  at (i2) {$i_2$};
    \node[right]  at (i3) {$i_3$};
    \node[right]  at (i4) {$i_4$};
    \node[right]  at (i5) {$i_5$};

    \draw (i1) -- (a) ;
    \draw (i2) -- (a) ;
    \draw (a) -- (b) ;
    \draw (b) -- (b) ;
    \draw (c) arc(0:360:0.5) --cycle;
    \draw (c) -- (d) ;
    \draw (d) -- (i3) ;
    \draw (d) -- (i4) ;
    \draw (d) -- (i5) ;
    
\end{tikzpicture}
+\tfrac{1}{8}
\begin{tikzpicture}[baseline={([yshift=-2.3pt]current bounding box.center)},scale=0.5]

    \coordinate (a) at (0,0);
    \coordinate (b) at (1,0);
    \coordinate (c) at (2,0);
    \coordinate (d) at (2.5,-0.5);

    \coordinate (i1) at (-0.5,1);
    \coordinate (i2) at (-0.5,-1);
    \coordinate (i3) at (2.5,-1.5);
    \coordinate (i4) at (3.5,-0.5);
    \coordinate (i5) at (3.5,1);

    \node[left]  at (i1) {$i_1$};
    \node[left]  at (i2) {$i_2$};
    \node[right]  at (i3) {$i_3$};
    \node[right]  at (i4) {$i_4$};
    \node[right]  at (i5) {$i_5$};

    \draw (i1) -- (a) ;
    \draw (i2) -- (a) ;
    \draw (c) -- (b) ;
    \draw (b) -- (b) ;
    \draw (b) arc(0:360:0.5) --cycle;
    \draw (c) -- (d) ;
    \draw (d) -- (i3) ;
    \draw (d) -- (i4) ;
    \draw (c) -- (i5) ;
    
\end{tikzpicture}
+\tfrac{1}{8}
\begin{tikzpicture}[baseline={([yshift=-2.3pt]current bounding box.center)},scale=0.5]

    \coordinate (a) at (0,0);
    \coordinate (b) at (1,0);
    \coordinate (c) at (2,0);
    \coordinate (d) at (2.5,-0.5);

    \coordinate (i1) at (-0.5,1);
    \coordinate (i2) at (-0.5,-1);
    \coordinate (i3) at (2.5,-1.5);
    \coordinate (i4) at (3.5,-0.5);
    \coordinate (i5) at (3.5,1);

    \node[left]  at (i1) {$i_1$};
    \node[left]  at (i2) {$i_2$};
    \node[right]  at (i3) {$i_3$};
    \node[right]  at (i4) {$i_4$};
    \node[right]  at (i5) {$i_5$};

    \draw (i1) -- (a) ;
    \draw (i2) -- (a) ;
    \draw (a) -- (b) ;
    \draw (b) -- (b) ;
    \draw (c) arc(0:360:0.5) --cycle;
    \draw (c) -- (d) ;
    \draw (d) -- (i3) ;
    \draw (d) -- (i4) ;
    \draw (c) -- (i5) ;
    
\end{tikzpicture}
\end{equation*}

\begin{equation*}
    \hspace{0.7cm}+\tfrac{1}{12}
    \begin{tikzpicture}[baseline={([yshift=-2.3pt]current bounding box.center)},scale=0.5]

    \coordinate (a) at (0,1);
    \coordinate (b) at (0,-1);
    \coordinate (c) at (1,0);
    \coordinate (d) at (2,0);

    \coordinate (i1) at (-0.5,1.5);
    \coordinate (i2) at (-0.5,-1.5);
    \coordinate (i3) at (3,-1);
    \coordinate (i4) at (3.5,0);
    \coordinate (i5) at (3,1);

    \node[left]  at (i1) {$i_1$};
    \node[left]  at (i2) {$i_2$};
    \node[right]  at (i3) {$i_3$};
    \node[right]  at (i4) {$i_4$};
    \node[right]  at (i5) {$i_5$};

    \draw (i1) -- (a) ;
    \draw (i2) -- (b) ;
    \draw (a) -- (b) ;
    \draw (b) -- (c) ;
    \draw (c) -- (a) ;
    \draw (c) -- (d) ;
    \draw (i3) -- (d) ;
    \draw (i4) -- (d) ;
    \draw (i5) -- (d) ;

    \end{tikzpicture}    
    +\tfrac{1}{4}
    \begin{tikzpicture}[baseline={([yshift=-2.3pt]current bounding box.center)},scale=0.5]

    \coordinate (a) at (0,0);
    \coordinate (b) at (1,0);
    \coordinate (c) at (3,0);
    \coordinate (d) at (2,-1);

    \coordinate (i1) at (-0.5,1);
    \coordinate (i2) at (-0.5,-1);
    \coordinate (i3) at (2,-2);
    \coordinate (i4) at (3.5,-1);
    \coordinate (i5) at (3.5,1);

    \node[left]  at (i1) {$i_1$};
    \node[left]  at (i2) {$i_2$};
    \node[right]  at (i3) {$i_3$};
    \node[right]  at (i4) {$i_4$};
    \node[right]  at (i5) {$i_5$};

    \draw (i1) -- (a) ;
    \draw (i2) -- (a) ;
    \draw (a) -- (b) ;
    \draw (b) -- (c) ;
    \draw (b) -- (d) ;
    \draw (c) -- (d);
    \draw (d) -- (i3) ;
    \draw (c) -- (i4) ;
    \draw (c) -- (i5) ;
    
    \end{tikzpicture}
    +\tfrac{1}{4}
    \begin{tikzpicture}[baseline={([yshift=-2.3pt]current bounding box.center)},scale=0.5]

    \coordinate (a) at (0,0);
    \coordinate (b) at (1,-1);
    \coordinate (c) at (2,0);
    \coordinate (d) at (2.5,0.5);

    \coordinate (i1) at (-1,0);
    \coordinate (i2) at (1,-1.5);
    \coordinate (i3) at (2.5,-0.5);
    \coordinate (i4) at (3,0.5);
    \coordinate (i5) at (2.5,1);

    \node[left]  at (i1) {$i_1$};
    \node[left]  at (i2) {$i_2$};
    \node[right]  at (i3) {$i_3$};
    \node[right]  at (i4) {$i_4$};
    \node[left]  at (i5) {$i_5$};

    \draw (i1) -- (a) ;
    \draw (a) -- (b) ;
    \draw (b) -- (c) ;
    \draw (c) -- (a) ;
    \draw (i2) -- (b) ;
    \draw (i3) -- (c) ;
    \draw (c) -- (d) ;
    \draw (i4) -- (d) ;
    \draw (i5) -- (d) ;

\end{tikzpicture}
\end{equation*}

\begin{equation}
    \hspace{-4.0cm}+\tfrac{1}{4}
    \begin{tikzpicture}[baseline={([yshift=-2.3pt]current bounding box.center)},scale=0.5]

    \coordinate (a) at (0,0);
    \coordinate (b) at (1,-1);
    \coordinate (c) at (2,0);
    \coordinate (d) at (1,1);

    \coordinate (i1) at (-1,0);
    \coordinate (i2) at (1,-2);
    \coordinate (i3) at (2.5,-1);
    \coordinate (i4) at (2.5,1);
    \coordinate (i5) at (1,2);

    \node[left]  at (i1) {$i_1$};
    \node[left]  at (i2) {$i_2$};
    \node[right]  at (i3) {$i_3$};
    \node[right]  at (i4) {$i_4$};
    \node[left]  at (i5) {$i_5$};

    \draw (a) -- (b) ;
    \draw (b) -- (c) ;
    \draw (c) -- (d) ;
    \draw (a) -- (d) ;
    \draw (i1) -- (a) ;
    \draw (i2) -- (b) ;
    \draw (i3) -- (c) ;
    \draw (i4) -- (c) ;
    \draw (i5) -- (d) ;

\end{tikzpicture}
+\textrm{perm}(1,2,3,4,5) \,.
\end{equation}
The three lines correspond respectively to bubble, triangle, and box loops. We will study the different groups separately. For the bubble diagrams, it is not hard to convince oneself that the leading contribution is $\mathcal{O}\big(1/(p_a\cdot p_5)^{2}\big)$, and comes from the first diagram when the soft leg is attached to the cubic coupling. Indeed, this is the only diagram where two propagators go on shell in the soft limit. This corresponds precisely to \cref{fig:Bubble}.

For the diagrams on the second line, the leading contribution comes from the first diagram when the soft leg is attached to a cubic vertex that is part of the loop, and the triangle loop is connected to the four-point vertex via a tree-level propagator. In this case, one factor of $1/(p_a\cdot p_5)$ comes from the triangle integral, while a second such factor comes from the intermediate propagator going on shell. Again, this corresponds precisely to \cref{fig:OneHardLine}. 

Finally, let us consider the box integral on the third line. We are summing over all permutations and must analyze the effect of each placement of the soft particle. The leading contribution is given by the diagram where the soft particle is attached to a cubic vertex that is opposite to the quartic vertex, and corresponds to \cref{fig:TwoHardLines}. Using the method of generalized unitarity \cite{Elvang_Huang_2015}, we find that 
\begin{equation}
 \begin{tikzpicture}[baseline={([yshift=-2.3pt]current bounding box.center)},scale=0.5]

    \coordinate (a) at (0,0);
    \coordinate (b) at (1,-1);
    \coordinate (c) at (2,0);
    \coordinate (d) at (1,1);

    \coordinate (i1) at (-1,0);
    \coordinate (i2) at (1,-2);
    \coordinate (i3) at (2.5,-1);
    \coordinate (i4) at (2.5,1);
    \coordinate (i5) at (1,2);

    \node[left]  at (i1) {$i_5$};
    \node[left]  at (i2) {$i_1$};
    \node[right]  at (i3) {$i_2$};
    \node[right]  at (i4) {$i_3$};
    \node[left]  at (i5) {$i_4$};

    \draw (a) -- (b) ;
    \draw (b) -- (c) ;
    \draw (c) -- (d) ;
    \draw (a) -- (d) ;
    \draw[soft] (i1) -- (a) ;
    \draw (i2) -- (b) ;
    \draw (i3) -- (c) ;
    \draw (i4) -- (c) ;
    \draw (i5) -- (d) ;

\end{tikzpicture}+\textrm{perm}(1,2,3,4)
= \sum_{a<b} I^{(1)}_{\Box} i V_{i_5}^{\;\;\; j_1 j_2} V_{i_a j_1}^{\quad j_a} V_{i_b j_2}^{\quad j_b} \cA_{4,\cdots j_a \cdots j_b}^{(0),V^0} + \mathcal{O}\left(\frac{1}{p_a\cdot p_5}\right)\,. 
\end{equation}
This exactly agrees with \cref{eq:KabTerm} as expected.  Note that diagrams where the soft leg is adjacent to the quartic vertex are subleading by one power of $p_5$.

By combining these contributions, we reproduce the leading quantum correction to the scalar soft theorem. We could have extended this analysis for more insertions of the cubic vertex, but we stop here for brevity.


\section{Outlook}\label{sec:Conclusion}

In this paper, we have investigated the scalar soft theorems in EFTs at one loop using the language of field-space geometry. For derivatively-coupled theories, the geometric soft theorem remains unchanged, while for theories with potential interactions, specifically $\phi^3$ and $\phi^4$ at one loop, the soft theorem receives quantum corrections. We derived the leading universal corrections to the soft theorem for general scalar EFTs for theories with potential interactions.

This work opens up many future avenues of research. We conjecture that the geometric soft theorem is valid at all-loop orders for derivatively-coupled theories.  The crucial missing ingredient to prove this result is a general argument that guarantees the absence of IR divergences for such theories to all-loop orders. If such an argument could be established,  the all-loop statement of the geometric soft theorem for derivatively-coupled theories would follow from the arguments presented in this paper combined with the original derivation \cite{Cheung:2021yog}.

There is still more to explore at one-loop order.  We provided the universal one-loop corrections to the soft theorems for theories with potential interactions.  It would be exciting to extend these results to subleading order in the soft expansion, or conversely to prove definitively that no universal subleading expressions exist.
Another immediate generalization of the analysis here is to study theories with massive scalars and theories where the scalars couple to fermions and gauge bosons. We expect that the \textit{generalized geometric soft theorem}, derived in ref.~\cite{Derda:2024jvo}, will hold at one-loop order in the absence of potential interactions (including Yukawa-type interactions), but a concrete demonstration of this conjecture has not been performed.  

Beyond the realm of single-particle soft theorems, several multi-particle soft theorems have been derived for tree-level scattering amplitudes \cite{Cheung:2021yog,Derda:2024jvo}. Whether any remnant of these results survives at the quantum level is an open question. Finally, one would hope that the deeper perspective this paper provides could lead to new understanding between the soft limit of scalar EFTs, asymptotic symmetries \cite{Campiglia:2017dpg,Campiglia:2017xkp,Henneaux:2018mgn,Biswas:2022lsj}, and the formalism of celestial holography \cite{Kapec:2022axw,Kapec:2022hih}. These connections could yield exciting new insights into the fundamental nature of EFTs.

\subsection*{Acknowledgments}

We thank Philipp Böer, Chia-Hsien Shen, and Gherardo Vita for insightful discussions and Julio Parra-Martinez for comments on the manuscript.
TC is supported by the U.S.~Department of Energy Grant No.~DE-SC0011640. FN and IP are partially supported by the Swiss National Science Foundation under contract 200020-213104 and through the National Center of Competence in Research SwissMAP. IP is supported by Scuola Normale Superiore. The research of IP is moreover supported by the ERC (NOTIMEFORCOSMO, 101126304). However, views and opinions expressed are those of the author(s) only and do not necessarily reflect those of the European Union or the European Research Council Executive Agency. Neither the European Union nor the granting authority can be held responsible for them.

\appendix
\section*{Appendix}
\addcontentsline{toc}{section}{\protect\numberline{}Appendix}%
\addtocontents{toc}{\protect\setcounter{tocdepth}{1}}

\section{Loop integrals}\label{app:Integrals}

Here we list the one-loop integrals that appear in the examples presented in \cref{app:Examples} above. A general one-loop $n$-point integral without any loop momentum dependence in the numerator takes the form 
\begin{align}
    I_{n} = \mu^{2\epsilon} \int \frac{\text{d}^{4-2\epsilon}\ell}{(2\pi)^{4-2\epsilon}} \frac{1}{\ell^2(\ell-p_1)^2(\ell-p_1-p_2)^2\cdots (\ell-p_1-p_2-\cdots -p_{n-1})^{2}}\,.
\end{align}
\subsection{Bubbles}
The massless bubble integral is scaleless and vanishes. The massive bubble integral is
\begin{align}
    \label{eq:IntegralBubble}
    I_2(p_1^2) = \frac{i}{16\pi^2}\bigg[\frac{1}{\epsilon}+2-\log\bigg(-\frac{p_1^2}{4\pi e^{-\gamma} \mu^2}\bigg)\bigg]+\mathcal{O}(\epsilon)\,.
\end{align}
The bubble integral with the propagators raised to arbitrary powers is
\begin{align}
    J_{2,\nu_1\nu_2}(p_1^2)&=\mu^{2\epsilon} \int \frac{\text{d}^{4-2\epsilon}\ell}{(2\pi)^{4-2\epsilon}} \frac{1}{\ell^{2\nu_1}(\ell-p_1)^{2\nu_2}} \notag\\[3pt]
   & =\frac{i}{16\pi^2}\frac{1}{(p_1^2)^{\nu_1+\nu_2-2}}\bigg(\frac{-p_1^2}{4\pi\mu^2}\bigg)^{-\epsilon}\notag\\[3pt]
   &\hspace{2cm}\times\frac{\Gamma(\nu_1 + \nu_2 -2 + \epsilon)\Gamma(2-\epsilon-\nu_1)\Gamma(2-\epsilon-\nu_2)}{\Gamma(\nu_1)\Gamma(\nu_2)\Gamma(4-2\epsilon-\nu_1-\nu_2)}\,.
\end{align}

\subsection{Triangles}
The one-mass triangle integral is
\begin{align}
    I^{1m}_{3}(p_1^2) = \frac{i}{16\pi^2} \frac{1}{p_1^2}\bigg[\frac{1}{\epsilon^2}-\frac{1}{\epsilon}\ln\bigg(\frac{-p_1^2}{4\pi e^{-\gamma}\mu^2}\bigg)+\frac{1}{2}\ln^2\bigg(\frac{-p_1^2}{4\pi e^{-\gamma}\mu^2}\bigg)-\frac{1}{2}\zeta_2 \bigg]+\mathcal{O}(\epsilon)\,.
\end{align}
In addition, we need the triangle integral with the propagators raised to arbitrary powers. We introduce the notation  $(\nu=\nu_1+\nu_2+\nu_3)$
\begin{align}
    &J^{1m}_{3,\nu_1\nu_2\nu_3}(p_1^2) = \mu^{2\epsilon} \int \frac{\text{d}^{4-2\epsilon}\ell}{(2\pi)^{4-2\epsilon}} \frac{1}{\ell^{2\nu_1}(\ell-p_1)^{2\nu_2}(\ell-p_1-p_2)^{2\nu_3}} \\[6pt]
    &=\frac{i}{16\pi^2}\frac{1}{(p_1^2)^{\nu-2}}\bigg(\!-\!\frac{p_1^2}{4\pi\mu^2}\bigg)^{-\epsilon}
    \frac{\Gamma(\nu-2+\epsilon)\Gamma(2-\epsilon - \nu_1 - \nu_3)\Gamma(2-\epsilon - \nu_2 - \nu_3)}{\Gamma(\nu_1)\Gamma(\nu_2)\Gamma(4-2\epsilon - \nu)}\notag\,.
\end{align}
It satisfies
\begin{equation}
    J^{1m}_{3,\nu_1\nu_2\nu_3}(p_1^2)=J^{1m}_{3,\nu_2\nu_1\nu_3}(p_1^2) \,.
\end{equation}
Some useful explicit cases are
\begin{align}
     J^{1m}_{3,121}(p_1^2) &=\frac{i}{16\pi^2} \frac{1}{(p_1^2)^2} \bigg[\frac{2}{\epsilon} -2\log\bigg(-\frac{p_1^2}{4\pi\mu^2e^{-\gamma}}\bigg)+\mathcal{O}(\epsilon)\bigg]\,,\\[8pt]
     J^{1m}_{3,112}(p_1^2) 
     &= \frac{i}{16\pi^2} \frac{1}{(p_1^2)^2} \bigg[-\frac{2}{\epsilon} +2+2\log\bigg(-\frac{p_1^2}{4\pi\mu^2e^{-\gamma}}\bigg)+\mathcal{O}(\epsilon)\bigg] \,.
\end{align}
The two-mass triangle integral is
\begin{equation}
    I_3^{2m}(p_2^2,p_3^2)=\frac{i}{16\pi^2}\frac{1}{p_3^2-p_2^2}\log\bigg(\frac{p_2^2}{p_3^2}\bigg)\bigg[\frac{1}{\epsilon}-\log\bigg(\frac{p_2^2\,p_3^2}{4\pi \mu^2e^{-2\gamma}}\bigg)+\mathcal{O(\epsilon)}\bigg]\,.
\end{equation}
The generalized version with propagators raised to arbitrary powers is
\begin{align}
    J^{2m}_{3,\nu_1\nu_2\nu_3}(p_2^2,p_3^2) &= \mu^{2\epsilon} \int \frac{\text{d}^{4-2\epsilon}\ell}{(2\pi)^{4-2\epsilon}} \frac{1}{\ell^{2\nu_1}(\ell-p_1)^{2\nu_2}(\ell-p_1-p_2)^{2\nu_3}}\notag\\[6pt]
    &=\frac{i}{16\pi^2}\frac{1}{(p_3^2)^{\nu-2}}\bigg(\frac{-p_3^2}{4\pi e^{-\gamma}\mu^2}\bigg)^{-\epsilon}\notag\\[6pt]
    &\hspace{12pt}\times\frac{\Gamma(\nu-2+\epsilon)\Gamma(2-\epsilon-\nu_1-\nu_2)\Gamma(2-\epsilon-\nu_3)}{\Gamma(4-2\epsilon-\nu)\Gamma(\nu_1+\nu_2)} \notag\\[6pt]
    &\hspace{12pt}\times {_2}F_1\big(\nu_2,\nu-2+\epsilon,\nu_1+\nu_2,1-p_2^2/p_3^2\big)\,.
\end{align}
It satisfies
\begin{equation}
    J^{2m}_{3,\nu_1\nu_2\nu_3}(p_2^2,p_3^2)=J^{2m}_{3,\nu_2\nu_1\nu_3}(p_3^2,p_2^2)\,.
\end{equation}
Some cases used above are
\begin{align}
     J^{2m}_{3,121}(p_2^2,p_3^2) &= \frac{i}{16\pi^2}\frac{2}{(p_2^2-p_3^2)^2}\bigg(\frac{p_2^2-p_3^2}{p_2^2}-\ln\bigg(\frac{p_2^2}{p_3^2}\bigg)\bigg) +\mathcal{O}(\epsilon)\,,\\[5pt]
     J^{2m}_{3,112}(p_2^2,p_3^2) &= \frac{i}{16\pi^2}\frac{1}{p_2^2\,p_3^2}\\[5pt]
     &\hspace{12pt}\times\bigg[-\frac{2}{\epsilon} -\frac{1}{p_3^2-p_2^2}\bigg(p_2^2\ln\bigg(\frac{-p_3^2}{4\pi e^{-\gamma}\mu^2}\bigg)-p_3^2\ln\bigg(\frac{-p_2^2}{4\pi e^{-\gamma}\mu^2}\bigg)\bigg)\bigg]+\mathcal{O}(\epsilon)\notag\,.
\end{align}
\subsection{Boxes}
The one-mass box integral is
\begin{align}
    I^{1m}_{4}\big((p_1+p_2)^2,(p_2+p_3)^2,m_4^2\big) &= \frac{i r_{\Gamma}}{(4\pi)^2} \frac{2}{s_{12}s_{23}}\left(\frac{-s_{12}}{4\pi\mu^2}\right)^{-\epsilon}\left(\frac{-s_{23}}{4\pi\mu^2}\right)^{-\epsilon}\left(\frac{-m_4^2}{4\pi\mu^2}\right)^{\epsilon}\nonumber \\[6pt] 
    &\hspace{12pt}\times
    \left[ \frac{1}{\epsilon^2} + \textrm{Li}_{2} \left(1-\frac{s_{12}}{m_4^2} \right) + \textrm{Li}_{2} \left(1-\frac{s_{23}}{m_4^2} \right) - \frac{\pi^2}{6}   \right]\,,\notag\\[2pt]
\end{align}
where
\begin{align}
    r_{\Gamma} = \frac{\Gamma(1+\epsilon)\Gamma^2(1-\epsilon)}{\Gamma(1-2\epsilon)}\,.
\end{align}
Expanding in $\epsilon$, we find
\begin{align}
    I^{1m}_{4}((p_1+p_2)^2,(p_2+p_3)^2,m_4^2) = \,&\frac{i}{16\pi^2}\frac{2}{s_{12}s_{23}}\bigg[\frac{1}{\epsilon^2}-\frac{1}{\epsilon} \log\bigg(-\frac{s_{12}}{4\pi\mu^2e^{-\gamma}}\frac{s_{23}}{m_4^2}\bigg)\notag\\[6pt]
    &\,+\frac{1}{2}\log^2\bigg(-\frac{s_{12}}{4\pi\mu^2e^{-\gamma}}\frac{s_{23}}{m_4^2}\bigg)- \frac{\pi^2}{4}\notag\\[6pt]
    &\,+\textrm{Li}_{2} \left(1-\frac{s_{12}}{m_4^2} \right) + \textrm{Li}_{2} \left(1-\frac{s_{23}}{m_4^2} \right)+\mathcal{O}(\epsilon)  \bigg]\,.\notag\\[2pt]
\end{align}

\clearpage

%
%
%


\end{spacing}

\begin{spacing}{1.09}
\addcontentsline{toc}{section}{\protect\numberline{}References}%
\bibliographystyle{JHEP}
\bibliography{softScalar.bib}

\providecommand{\href}[2]{#2}\begingroup\raggedright\begin{thebibliography}{10}

\bibitem{Weinberg:1965nx}
S.~Weinberg, \emph{{Infrared photons and gravitons}}, \href{https://doi.org/10.1103/PhysRev.140.B516}{\emph{Phys. Rev.} {\bfseries 140} (1965) B516}.

\bibitem{Adler:1964um}
S.L.~Adler, \emph{{Consistency conditions on the strong interactions implied by a partially conserved axial vector current}}, \href{https://doi.org/10.1103/PhysRev.137.B1022}{\emph{Phys. Rev.} {\bfseries 137} (1965) B1022}.

\bibitem{Green:2022slj}
D.~Green, Y.~Huang and C.-H.~Shen, \emph{{Inflationary Adler conditions}}, \href{https://doi.org/10.1103/PhysRevD.107.043534}{\emph{Phys. Rev. D} {\bfseries 107} (2023) 043534} [\href{https://arxiv.org/abs/2208.14544}{{\ttfamily 2208.14544}}].

\bibitem{Cheung:2023qwn}
C.~Cheung, M.~Derda, A.~Helset and J.~Parra-Martinez, \emph{{Soft phonon theorems}}, \href{https://doi.org/10.1007/JHEP08(2023)103}{\emph{JHEP} {\bfseries 08} (2023) 103} [\href{https://arxiv.org/abs/2301.11363}{{\ttfamily 2301.11363}}].

\bibitem{Weinberg:1966kf}
S.~Weinberg, \emph{{Pion scattering lengths}}, \href{https://doi.org/10.1103/PhysRevLett.17.616}{\emph{Phys. Rev. Lett.} {\bfseries 17} (1966) 616}.

\bibitem{Pasterski:2021raf}
S.~Pasterski, M.~Pate and A.-M.~Raclariu, \emph{{Celestial Holography}},  in \emph{{Snowmass 2021}}, 11, 2021 [\href{https://arxiv.org/abs/2111.11392}{{\ttfamily 2111.11392}}].

\bibitem{Yennie:1961ad}
D.R.~Yennie, S.C.~Frautschi and H.~Suura, \emph{{The infrared divergence phenomena and high-energy processes}}, \href{https://doi.org/10.1016/0003-4916(61)90151-8}{\emph{Annals Phys.} {\bfseries 13} (1961) 379}.

\bibitem{Grammer:1973db}
G.~Grammer, Jr. and D.R.~Yennie, \emph{{Improved treatment for the infrared divergence problem in quantum electrodynamics}}, \href{https://doi.org/10.1103/PhysRevD.8.4332}{\emph{Phys. Rev. D} {\bfseries 8} (1973) 4332}.

\bibitem{Catani:2000pi}
S.~Catani and M.~Grazzini, \emph{{The soft gluon current at one loop order}}, \href{https://doi.org/10.1016/S0550-3213(00)00572-1}{\emph{Nucl. Phys. B} {\bfseries 591} (2000) 435} [\href{https://arxiv.org/abs/hep-ph/0007142}{{\ttfamily hep-ph/0007142}}].

\bibitem{Larkoski:2014bxa}
A.J.~Larkoski, D.~Neill and I.W.~Stewart, \emph{{Soft Theorems from Effective Field Theory}}, \href{https://doi.org/10.1007/JHEP06(2015)077}{\emph{JHEP} {\bfseries 06} (2015) 077} [\href{https://arxiv.org/abs/1412.3108}{{\ttfamily 1412.3108}}].

\bibitem{Bern:2014oka}
Z.~Bern, S.~Davies and J.~Nohle, \emph{{On Loop Corrections to Subleading Soft Behavior of Gluons and Gravitons}}, \href{https://doi.org/10.1103/PhysRevD.90.085015}{\emph{Phys. Rev. D} {\bfseries 90} (2014) 085015} [\href{https://arxiv.org/abs/1405.1015}{{\ttfamily 1405.1015}}].

\bibitem{Beneke:2022pue}
M.~Beneke, P.~Hager and R.~Szafron, \emph{{Soft-Collinear Gravity and Soft Theorems}},  \href{https://arxiv.org/abs/2210.09336}{{\ttfamily 2210.09336}}.

\bibitem{Cheung:2021yog}
C.~Cheung, A.~Helset and J.~Parra-Martinez, \emph{{Geometric soft theorems}}, \href{https://doi.org/10.1007/JHEP04(2022)011}{\emph{JHEP} {\bfseries 04} (2022) 011} [\href{https://arxiv.org/abs/2111.03045}{{\ttfamily 2111.03045}}].

\bibitem{Alonso:2015fsp}
R.~Alonso, E.E.~Jenkins and A.V.~Manohar, \emph{{A Geometric Formulation of Higgs Effective Field Theory: Measuring the Curvature of Scalar Field Space}}, \href{https://doi.org/10.1016/j.physletb.2016.01.041}{\emph{Phys. Lett. B} {\bfseries 754} (2016) 335} [\href{https://arxiv.org/abs/1511.00724}{{\ttfamily 1511.00724}}].

\bibitem{Alonso:2016oah}
R.~Alonso, E.E.~Jenkins and A.V.~Manohar, \emph{{Geometry of the Scalar Sector}}, \href{https://doi.org/10.1007/JHEP08(2016)101}{\emph{JHEP} {\bfseries 08} (2016) 101} [\href{https://arxiv.org/abs/1605.03602}{{\ttfamily 1605.03602}}].

\bibitem{Helset:2018fgq}
A.~Helset, M.~Paraskevas and M.~Trott, \emph{{Gauge fixing the Standard Model Effective Field Theory}}, \href{https://doi.org/10.1103/PhysRevLett.120.251801}{\emph{Phys. Rev. Lett.} {\bfseries 120} (2018) 251801} [\href{https://arxiv.org/abs/1803.08001}{{\ttfamily 1803.08001}}].

\bibitem{Corbett:2019cwl}
T.~Corbett, A.~Helset and M.~Trott, \emph{{Ward Identities for the Standard Model Effective Field Theory}}, \href{https://doi.org/10.1103/PhysRevD.101.013005}{\emph{Phys. Rev. D} {\bfseries 101} (2020) 013005} [\href{https://arxiv.org/abs/1909.08470}{{\ttfamily 1909.08470}}].

\bibitem{Helset:2020yio}
A.~Helset, A.~Martin and M.~Trott, \emph{{The Geometric Standard Model Effective Field Theory}}, \href{https://doi.org/10.1007/JHEP03(2020)163}{\emph{JHEP} {\bfseries 03} (2020) 163} [\href{https://arxiv.org/abs/2001.01453}{{\ttfamily 2001.01453}}].

\bibitem{Hays:2020scx}
C.~Hays, A.~Helset, A.~Martin and M.~Trott, \emph{{Exact SMEFT formulation and expansion to $\mathcal{O}(v^4/\Lambda^4)$}}, \href{https://doi.org/10.1007/JHEP11(2020)087}{\emph{JHEP} {\bfseries 11} (2020) 087} [\href{https://arxiv.org/abs/2007.00565}{{\ttfamily 2007.00565}}].

\bibitem{Cohen:2020xca}
T.~Cohen, N.~Craig, X.~Lu and D.~Sutherland, \emph{{Is SMEFT Enough?}}, \href{https://doi.org/10.1007/JHEP03(2021)237}{\emph{JHEP} {\bfseries 03} (2021) 237} [\href{https://arxiv.org/abs/2008.08597}{{\ttfamily 2008.08597}}].

\bibitem{Corbett:2021eux}
T.~Corbett, A.~Helset, A.~Martin and M.~Trott, \emph{{EWPD in the SMEFT to dimension eight}}, \href{https://doi.org/10.1007/JHEP06(2021)076}{\emph{JHEP} {\bfseries 06} (2021) 076} [\href{https://arxiv.org/abs/2102.02819}{{\ttfamily 2102.02819}}].

\bibitem{Cohen:2021ucp}
T.~Cohen, N.~Craig, X.~Lu and D.~Sutherland, \emph{{Unitarity violation and the geometry of Higgs EFTs}}, \href{https://doi.org/10.1007/JHEP12(2021)003}{\emph{JHEP} {\bfseries 12} (2021) 003} [\href{https://arxiv.org/abs/2108.03240}{{\ttfamily 2108.03240}}].

\bibitem{Alonso:2021rac}
R.~Alonso and M.~West, \emph{{Roads to the Standard Model}}, \href{https://doi.org/10.1103/PhysRevD.105.096028}{\emph{Phys. Rev. D} {\bfseries 105} (2022) 096028} [\href{https://arxiv.org/abs/2109.13290}{{\ttfamily 2109.13290}}].

\bibitem{Helset:2022pde}
A.~Helset, E.E.~Jenkins and A.V.~Manohar, \emph{{Renormalization of the Standard Model Effective Field Theory from geometry}}, \href{https://doi.org/10.1007/JHEP02(2023)063}{\emph{JHEP} {\bfseries 02} (2023) 063} [\href{https://arxiv.org/abs/2212.03253}{{\ttfamily 2212.03253}}].

\bibitem{Alonso:2022ffe}
R.~Alonso and M.~West, \emph{{On the effective action for scalars in a general manifold to any loop order}}, \href{https://doi.org/10.1016/j.physletb.2023.137937}{\emph{Phys. Lett. B} {\bfseries 841} (2023) 137937} [\href{https://arxiv.org/abs/2207.02050}{{\ttfamily 2207.02050}}].

\bibitem{Assi:2023zid}
B.~Assi, A.~Helset, A.V.~Manohar, J.~Pag\`es and C.-H.~Shen, \emph{{Fermion geometry and the renormalization of the Standard Model Effective Field Theory}}, \href{https://doi.org/10.1007/JHEP11(2023)201}{\emph{JHEP} {\bfseries 11} (2023) 201} [\href{https://arxiv.org/abs/2307.03187}{{\ttfamily 2307.03187}}].

\bibitem{Jenkins:2023bls}
E.E.~Jenkins, A.V.~Manohar, L.~Naterop and J.~Pag\`es, \emph{{Two loop renormalization of scalar theories using a geometric approach}}, \href{https://doi.org/10.1007/JHEP02(2024)131}{\emph{JHEP} {\bfseries 02} (2024) 131} [\href{https://arxiv.org/abs/2310.19883}{{\ttfamily 2310.19883}}].

\bibitem{Alonso:2023upf}
R.~Alonso, \emph{{A primer on Higgs Effective Field Theory with Geometry}},  \href{https://arxiv.org/abs/2307.14301}{{\ttfamily 2307.14301}}.

\bibitem{Li:2024ciy}
X.-X.~Li, X.~Lu and Z.~Zhang, \emph{{The Geometric Universal One-Loop Effective Action}},  \href{https://arxiv.org/abs/2411.04173}{{\ttfamily 2411.04173}}.

\bibitem{Helset:2022tlf}
A.~Helset, E.E.~Jenkins and A.V.~Manohar, \emph{{Geometry in scattering amplitudes}}, \href{https://doi.org/10.1103/PhysRevD.106.116018}{\emph{Phys. Rev. D} {\bfseries 106} (2022) 116018} [\href{https://arxiv.org/abs/2210.08000}{{\ttfamily 2210.08000}}].

\bibitem{Helset:2024vle}
A.~Helset, \emph{{Color-kinematics duality for nonlinear sigma models with nonsymmetric cosets}}, \href{https://doi.org/10.1103/PhysRevD.110.L101701}{\emph{Phys. Rev. D} {\bfseries 110} (2024) L101701} [\href{https://arxiv.org/abs/2406.10955}{{\ttfamily 2406.10955}}].

\bibitem{Lee:2024xqa}
Y.-T.~Lee, \emph{{Field Space Geometry and Nonlinear Supersymmetry}},  \href{https://arxiv.org/abs/2410.21395}{{\ttfamily 2410.21395}}.

\bibitem{Finn:2019aip}
K.~Finn, S.~Karamitsos and A.~Pilaftsis, \emph{{Frame Covariance in Quantum Gravity}}, \href{https://doi.org/10.1103/PhysRevD.102.045014}{\emph{Phys. Rev. D} {\bfseries 102} (2020) 045014} [\href{https://arxiv.org/abs/1910.06661}{{\ttfamily 1910.06661}}].

\bibitem{Cheung:2022vnd}
C.~Cheung, A.~Helset and J.~Parra-Martinez, \emph{{Geometry-kinematics duality}}, \href{https://doi.org/10.1103/PhysRevD.106.045016}{\emph{Phys. Rev. D} {\bfseries 106} (2022) 045016} [\href{https://arxiv.org/abs/2202.06972}{{\ttfamily 2202.06972}}].

\bibitem{Cohen:2022uuw}
T.~Cohen, N.~Craig, X.~Lu and D.~Sutherland, \emph{{On-Shell Covariance of Quantum Field Theory Amplitudes}}, \href{https://doi.org/10.1103/PhysRevLett.130.041603}{\emph{Phys. Rev. Lett.} {\bfseries 130} (2023) 041603} [\href{https://arxiv.org/abs/2202.06965}{{\ttfamily 2202.06965}}].

\bibitem{Cohen:2023ekv}
T.~Cohen, X.~Lu and D.~Sutherland, \emph{{On amplitudes and field redefinitions}}, \href{https://doi.org/10.1007/JHEP06(2024)149}{\emph{JHEP} {\bfseries 06} (2024) 149} [\href{https://arxiv.org/abs/2312.06748}{{\ttfamily 2312.06748}}].

\bibitem{Craig:2023wni}
N.~Craig, Y.-T.~Lee, X.~Lu and D.~Sutherland, \emph{{Effective field theories as Lagrange spaces}}, \href{https://doi.org/10.1007/JHEP11(2023)069}{\emph{JHEP} {\bfseries 11} (2023) 069} [\href{https://arxiv.org/abs/2305.09722}{{\ttfamily 2305.09722}}].

\bibitem{Craig:2023hhp}
N.~Craig and Y.-T.~Lee, \emph{{Effective Field Theories on the Jet Bundle}}, \href{https://doi.org/10.1103/PhysRevLett.132.061602}{\emph{Phys. Rev. Lett.} {\bfseries 132} (2024) 061602} [\href{https://arxiv.org/abs/2307.15742}{{\ttfamily 2307.15742}}].

\bibitem{Alminawi:2023qtf}
M.~Alminawi, I.~Brivio and J.~Davighi, \emph{{Jet bundle geometry of scalar field theories}}, \href{https://doi.org/10.1088/1751-8121/ad72bb}{\emph{J. Phys. A} {\bfseries 57} (2024) 435401} [\href{https://arxiv.org/abs/2308.00017}{{\ttfamily 2308.00017}}].

\bibitem{Cohen:2024bml}
T.~Cohen, X.~Lu and Z.~Zhang, \emph{{What is the geometry of effective field theories?}}, \href{https://doi.org/10.1103/PhysRevD.111.085012}{\emph{Phys. Rev. D} {\bfseries 111} (2025) 085012} [\href{https://arxiv.org/abs/2410.21378}{{\ttfamily 2410.21378}}].

\bibitem{Aigner:2025xyt}
P.~Aigner, L.~Bellafronte, E.~Gendy, D.~Haslehner and A.~Weiler, \emph{{Renormalising the Field-Space Geometry}},  \href{https://arxiv.org/abs/2503.09785}{{\ttfamily 2503.09785}}.

\bibitem{Volkov:1973vd}
D.V.~Volkov, \emph{{Phenomenological Lagrangians}}, {\emph{Fiz. Elem. Chast. Atom. Yadra} {\bfseries 4} (1973) 3}.

\bibitem{Derda:2024jvo}
M.~Derda, A.~Helset and J.~Parra-Martinez, \emph{{Soft scalars in effective field theory}}, \href{https://doi.org/10.1007/JHEP06(2024)133}{\emph{JHEP} {\bfseries 06} (2024) 133} [\href{https://arxiv.org/abs/2403.12142}{{\ttfamily 2403.12142}}].

\bibitem{Bern:1995ix}
Z.~Bern and G.~Chalmers, \emph{{Factorization in one loop gauge theory}}, \href{https://doi.org/10.1016/0550-3213(95)00226-I}{\emph{Nucl. Phys. B} {\bfseries 447} (1995) 465} [\href{https://arxiv.org/abs/hep-ph/9503236}{{\ttfamily hep-ph/9503236}}].

\bibitem{Bern:1992em}
Z.~Bern, L.J.~Dixon and D.A.~Kosower, \emph{{Dimensionally regulated one loop integrals}}, \href{https://doi.org/10.1016/0370-2693(93)90400-C}{\emph{Phys. Lett. B} {\bfseries 302} (1993) 299} [\href{https://arxiv.org/abs/hep-ph/9212308}{{\ttfamily hep-ph/9212308}}].

\bibitem{Charap:1970xj}
J.M.~Charap, \emph{{Closed-loop calculations using a chiral-invariant lagrangian}}, \href{https://doi.org/10.1103/PhysRevD.3.1998}{\emph{Phys. Rev. D} {\bfseries 2} (1970) 1554}.

\bibitem{Bartsch:2024ofb}
C.~Bartsch, K.~Kampf, J.~Novotny and J.~Trnka, \emph{{All-loop soft theorem for pions}}, \href{https://doi.org/10.1103/PhysRevD.110.045009}{\emph{Phys. Rev. D} {\bfseries 110} (2024) 045009} [\href{https://arxiv.org/abs/2401.04731}{{\ttfamily 2401.04731}}].

\bibitem{Chisholm:1961tha}
J.S.R.~Chisholm, \emph{{Change of variables in quantum field theories}}, \href{https://doi.org/10.1016/0029-5582(61)90106-7}{\emph{Nucl. Phys.} {\bfseries 26} (1961) 469}.

\bibitem{Kamefuchi:1961sb}
S.~Kamefuchi, L.~O'Raifeartaigh and A.~Salam, \emph{{Change of variables and equivalence theorems in quantum field theories}}, \href{https://doi.org/10.1016/0029-5582(61)90056-6}{\emph{Nucl. Phys.} {\bfseries 28} (1961) 529}.

\bibitem{tHooft:1973wag}
G.~'t~Hooft and M.J.G.~Veltman, \emph{{DIAGRAMMAR}}, \href{https://doi.org/10.1007/978-1-4684-2826-1_5}{\emph{NATO Sci. Ser. B} {\bfseries 4} (1974) 177}.

\bibitem{Coleman:1969sm}
S.R.~Coleman, J.~Wess and B.~Zumino, \emph{{Structure of phenomenological Lagrangians. 1.}}, \href{https://doi.org/10.1103/PhysRev.177.2239}{\emph{Phys. Rev.} {\bfseries 177} (1969) 2239}.

\bibitem{Callan:1969sn}
C.G.~Callan, Jr., S.R.~Coleman, J.~Wess and B.~Zumino, \emph{{Structure of phenomenological Lagrangians. 2.}}, \href{https://doi.org/10.1103/PhysRev.177.2247}{\emph{Phys. Rev.} {\bfseries 177} (1969) 2247}.

\bibitem{Deans:1978wn}
W.S.~Deans and J.A.~Dixon, \emph{{Theory of Gauge Invariant Operators: Their Renormalization and S Matrix Elements}}, \href{https://doi.org/10.1103/PhysRevD.18.1113}{\emph{Phys. Rev. D} {\bfseries 18} (1978) 1113}.

\bibitem{Politzer:1980me}
H.D.~Politzer, \emph{{Power Corrections at Short Distances}}, \href{https://doi.org/10.1016/0550-3213(80)90172-8}{\emph{Nucl. Phys. B} {\bfseries 172} (1980) 349}.

\bibitem{Arzt:1993gz}
C.~Arzt, \emph{{Reduced effective Lagrangians}}, \href{https://doi.org/10.1016/0370-2693(94)01419-D}{\emph{Phys. Lett. B} {\bfseries 342} (1995) 189} [\href{https://arxiv.org/abs/hep-ph/9304230}{{\ttfamily hep-ph/9304230}}].

\bibitem{Passarino:2016saj}
G.~Passarino, \emph{{Field reparametrization in effective field theories}}, \href{https://doi.org/10.1140/epjp/i2017-11291-5}{\emph{Eur. Phys. J. Plus} {\bfseries 132} (2017) 16} [\href{https://arxiv.org/abs/1610.09618}{{\ttfamily 1610.09618}}].

\bibitem{Criado:2018sdb}
J.C.~Criado and M.~P\'erez-Victoria, \emph{{Field redefinitions in effective theories at higher orders}}, \href{https://doi.org/10.1007/JHEP03(2019)038}{\emph{JHEP} {\bfseries 03} (2019) 038} [\href{https://arxiv.org/abs/1811.09413}{{\ttfamily 1811.09413}}].

\bibitem{Cohen:2024fak}
T.~Cohen, M.~Forslund and A.~Helset, \emph{{Field Redefinitions Can Be Nonlocal}},  \href{https://arxiv.org/abs/2412.12247}{{\ttfamily 2412.12247}}.

\bibitem{Bern:1994zx}
Z.~Bern, L.J.~Dixon, D.C.~Dunbar and D.A.~Kosower, \emph{{One loop n point gauge theory amplitudes, unitarity and collinear limits}}, \href{https://doi.org/10.1016/0550-3213(94)90179-1}{\emph{Nucl. Phys. B} {\bfseries 425} (1994) 217} [\href{https://arxiv.org/abs/hep-ph/9403226}{{\ttfamily hep-ph/9403226}}].

\bibitem{Bern:1994cg}
Z.~Bern, L.J.~Dixon, D.C.~Dunbar and D.A.~Kosower, \emph{{Fusing gauge theory tree amplitudes into loop amplitudes}}, \href{https://doi.org/10.1016/0550-3213(94)00488-Z}{\emph{Nucl. Phys. B} {\bfseries 435} (1995) 59} [\href{https://arxiv.org/abs/hep-ph/9409265}{{\ttfamily hep-ph/9409265}}].

\bibitem{Cachazo:2014dia}
F.~Cachazo and E.Y.~Yuan, \emph{{Are Soft Theorems Renormalized?}},  \href{https://arxiv.org/abs/1405.3413}{{\ttfamily 1405.3413}}.

\bibitem{Beneke:1997zp}
M.~Beneke and V.A.~Smirnov, \emph{{Asymptotic expansion of Feynman integrals near threshold}}, \href{https://doi.org/10.1016/S0550-3213(98)00138-2}{\emph{Nucl. Phys. B} {\bfseries 522} (1998) 321} [\href{https://arxiv.org/abs/hep-ph/9711391}{{\ttfamily hep-ph/9711391}}].

\bibitem{Elvang_Huang_2015}
H.~Elvang and Y.-t.~Huang, \emph{Scattering Amplitudes in Gauge Theory and Gravity}, Cambridge University Press (2015).

\bibitem{Campiglia:2017dpg}
M.~Campiglia, L.~Coito and S.~Mizera, \emph{{Can scalars have asymptotic symmetries?}}, \href{https://doi.org/10.1103/PhysRevD.97.046002}{\emph{Phys. Rev. D} {\bfseries 97} (2018) 046002} [\href{https://arxiv.org/abs/1703.07885}{{\ttfamily 1703.07885}}].

\bibitem{Campiglia:2017xkp}
M.~Campiglia and L.~Coito, \emph{{Asymptotic charges from soft scalars in even dimensions}}, \href{https://doi.org/10.1103/PhysRevD.97.066009}{\emph{Phys. Rev. D} {\bfseries 97} (2018) 066009} [\href{https://arxiv.org/abs/1711.05773}{{\ttfamily 1711.05773}}].

\bibitem{Henneaux:2018mgn}
M.~Henneaux and C.~Troessaert, \emph{{Asymptotic structure of a massless scalar field and its dual two-form field at spatial infinity}}, \href{https://doi.org/10.1007/JHEP05(2019)147}{\emph{JHEP} {\bfseries 05} (2019) 147} [\href{https://arxiv.org/abs/1812.07445}{{\ttfamily 1812.07445}}].

\bibitem{Biswas:2022lsj}
S.~Biswas and G.W.~Semenoff, \emph{{Soft scalars do not decouple}}, \href{https://doi.org/10.1103/PhysRevD.106.105023}{\emph{Phys. Rev. D} {\bfseries 106} (2022) 105023} [\href{https://arxiv.org/abs/2208.05023}{{\ttfamily 2208.05023}}].

\bibitem{Kapec:2022axw}
D.~Kapec, Y.T.A.~Law and S.A.~Narayanan, \emph{{Soft scalars and the geometry of the space of celestial conformal field theories}}, \href{https://doi.org/10.1103/PhysRevD.107.046024}{\emph{Phys. Rev. D} {\bfseries 107} (2023) 046024} [\href{https://arxiv.org/abs/2205.10935}{{\ttfamily 2205.10935}}].

\bibitem{Kapec:2022hih}
D.~Kapec, \emph{{Soft particles and infinite-dimensional geometry}}, \href{https://doi.org/10.1088/1361-6382/ad0514}{\emph{Class. Quant. Grav.} {\bfseries 41} (2024) 015001} [\href{https://arxiv.org/abs/2210.00606}{{\ttfamily 2210.00606}}].

\end{thebibliography}\endgroup
\end{spacing}

\end{document}